\documentclass[11pt, referee] {aa}

\usepackage{graphicx, subfig}
\usepackage{natbib,twoopt, units}
\usepackage[breaklinks=true]{hyperref} %% to avoid \citeads line fills 
\bibpunct{(}{)}{;}{a}{}{,} %% natbib format like A&A and ApJ 
\newcommandtwoopt{\citeads}[3][][]{\href{http://adsabs.harvard.edu/abs/}% 
{\citealp[#1][#2]{#3}}} 
\newcommandtwoopt{\citepads}[3][][]{\href{http://adsabs.harvard.edu/abs/#3}% 
{\citep[#1][#2]{#3}}} 
\newcommandtwoopt{\citetads}[3][][]{\href{http://adsabs.harvard.edu/abs/#3}% 
{\citet[#1][#2]{#3}}} 
\newcommandtwoopt{\citeyearads}[3][][]% 
{\href{http://adsabs.harvard.edu/abs/#3}{\citeyear[#1][#2]{#3}}}
\usepackage{hyperref}
\hypersetup{
  pdfinfo={
  pdfproducer={},
  Title={},
  Subject={},
  Author={},
  Creator={},
  Producer={},
  }
}

\title{From stellar nebula to planets: The refractory components}

\authorrunning{Thiabaud et al.}
\titlerunning{The refractory components}

\author{Amaury~Thiabaud \inst{1,2}, Ulysse~Marboeuf\inst{1,2}, Yann Alibert \inst{1,2,3}, Nahuel Cabral\inst{1,2}, Ingo Leya\inst{1,2} \& Klaus Mezger \inst{1,4} }
\institute{$^1$Center for Space and Habitability, Universit\"{a}t Bern, CH-3012 Bern, Switzerland.\\   
Email: amaury.thiabaud@csh.unibe.ch\\
$^2$Physikalisches Institut, Universit\"{a}t Bern, CH-3012 Bern, Switzerland\\
$^3$Observatoire de Besan\c{c}on, 41 Avenue de l'Observatoire, 25000 Besan\c{c}on, France\\
$^4$Institut f\"{u}r Geologie, Universit\"{a}t Bern, CH-3012 Bern, Switzerland}

\date{Received ??; accepted ??}

\begin{document}

	\abstract
   	{To date calculations of planet formation have mainly focused on dynamics, and only a few have considered the chemical composition of refractory elements and compounds in the planetary bodies. While many studies have been concentrating on the chemical composition of volatile compounds (such as H$_{2}$O, CO, CO$_{2}$) incorporated in planets, only a few have considered the refractory materials as well, although they are of great importance for the formation of rocky planets.} 
   	{We computed the abundance of refractory elements in planetary bodies formed in stellar systems with a solar chemical composition by combining models of chemical composition and planet formation.  We also consider the formation of refractory organic compounds, which have been ignored in previous studies on this topic. }
   	{We used the commercial software package HSC Chemistry to compute the condensation sequence and chemical composition of refractory minerals incorporated into planets. The problem of refractory organic material is approached with two distinct model calculations: the first considers that the fraction of atoms used in the formation of organic compounds is removed from the system (i.e., organic compounds are formed in the gas phase and are non-reactive); and the second assumes that organic compounds are formed by the reaction between different compounds that had previously condensed from the gas phase.}
   	{Results show that refractory material represents more than 50 wt \% of the mass of solids accreted by the simulated planets with up to 30 wt \% of the total mass composed of refractory organic compounds. Carbide and silicate abundances are consistent with C/O and Mg/Si elemental ratios of 0.5 and 1.02 for the Sun. Less than 1 wt \% of carbides are present in the planets, and pyroxene and olivine are formed in similar quantities. The model predicts planets  that are similar in composition to those of the solar system. Starting from a common initial nebula composition, it also shows that a wide variety of chemically different planets can form, which means that the differences in planetary compositions are due to differences in the planetary formation process.}
   	 {We show that a model in which refractory organic material is absent from the system is more compatible with observations. The use of a planet formation model is essential to form a wide diversity of planets in a consistent way.}

   
   	\maketitle
	
	\section{Introduction}
	        The chemical and dynamical processes involved in planet formation are still not fully understood. Current models describe the formation of planets and planetesimals by accretion of solid material and interactions among rocky cores, planetesimals, dust, and gas. This so-called core-accretion model has been developed by \cite{Pollack1996} and is included in the work of \citet{Alibert2005d, Mordasini2009, Mordasini2009a, Alibert2010, Mordasini2012a, Fortier2013}; \cite{Alibert2013} (citing only papers related to the model used here); \cite{Hubickyj2005}; and \cite{Lissauer2009}. Many attempts have been made to simulate the formation of our solar system, but these calculations have failed to reproduce key features of the Solar System \citep[see][for a review]{Chambers2004}.\\
		
		To date, calculations of planetary compositions are rare, although interest in such models has increased with the discovery of exoplanets and the search for habitable planets that could harbor life. The need to know the composition of planets thus grows strongly as the planetary composition is directly linked to the planetary habitability. So far, only atmospheres of planets can be studied directly thanks to spectroscopic data from observations of transiting planets. These studies suggest that most planets are different from the Earth in terms of their abundances of gases \citep[see][]{Madhusudhan2012}. Until now, only calculations of the planetary composition involving volatile species have been carried out \citep[see e.g.][]{Mousis2005a, Marboeuf2008, Mousis2009, Johnson2012} and combined with strong constraints from observations of exoplanet atmospheres and models of ice formation. A major goal of these models has been the evaluation of the possible presence of liquid water on planetary surfaces \citep{Johnson2012}. \\
		
		So far, only \citet{Bond2010, Bond2010a}, \citet{Johnson2012}, and \cite{Elser2012} have also examined the concentration of refractory elements in planets in an attempt to model the formation of planets that are chemically similar to the rocky planets of our Solar System or as proxies for predicting abundances of exoplanets. These calculations, which are based on condensation sequences under equilibrium conditions, try to combine dynamic and chemical models more rigorously than studies focusing on volatile compounds. They succeeded at correlating the composition of the host star with the composition of planets via the two element ratios C/O and Mg/Si. \cite{DelgadoMena2010}  showed that these two ratios are proxies for the distribution of carbides and silicates in planetary systems. \\
		
		In the present study, the abundance of refractory compounds incorporated into planets is quantified using the commercial software package HSC Chemistry (v. 7.1) and the model for planet formation of \cite{Alibert2005d} and \cite{Alibert2013}. In Sect. \ref{dyn_mod}, we summarize the main assumptions of this planet formation model. Section  \ref{chem_mod} is dedicated to the description of the chemical model used to determine refractory and volatile molecule abundances in the formed planets. Results and discussions are presented in Sect. \ref{results} and Sect.\ref{discussions} respectively. Section \ref{ccl} is devoted to conclusions.

	\section{Planet formation model} \label{dyn_mod}
		
		The calculations are based on the planet formation model of \citet{Alibert2005d, Mordasini2012, Fortier2013}  and \cite{Alibert2013} and are used to compute the growth process and migration of planets. The planet formation model is based on the so-called "core accretion model", in which a solid core is first formed by accretion of solid planetesimals, which themselves were formed by coagulation and sedimentation of small dust grains. This core will eventually be massive enough to gravitationally bind some of the nebular gas, thus surrounding itself with a tenuous envelope. %The growth of core and envelope follow
		
		The model of \cite{Alibert2005d} is structured around three modules. These modules calculate the structure and the evolution of a non-irradiated disc, including planet migration, interaction of planetesimals with the planet's atmosphere, and the planet's structure and evolution. The structure and evolution of the proto-planetary disc is computed by determining the vertical structure of the disc for each distance to the central star, which is used to compute the radial evolution as a function of viscosity, photo-evaporation rate, and mass accretion rate. The vertical structure of the disc is computed by solving the hydrostatic equilibrium equation, the energy conservation equation, and the diffusion equation for the radiative flux \citep[for details, see][]{Fortier2013}, while the evolution of the disc is computed through solving the diffusion equation. It is assumed in this model that planetesimals are not subject to radial drift. \\
		
		%The aforementioned 
		This model allows us to determine the evolution of the temperature T(r), pressure P(r), and surface density $\Sigma$(T,P,r) of the disc for different distances to the star (ranging from 0.04 to 30 AU) and for different viscosities $\nu$(r, z) and masses of the disc. The results of \cite{Alibert2013} are used to simulate 500 planetary systems, whose composition is modeled. %???.
		 Each system initially contains planetesimals and ten planetary embryos of lunar mass. After the dynamical evolution of the 500 systems, 4900 planets have been formed. 		
	
	\section{Chemical model} \label{chem_mod}
		\subsection{Refractory elements}
			\subsubsection{Inorganic compounds}
			
			The abundance of refractory minerals in planetesimals is supposed to be ruled by equilibrium condensation of the material from the primordial stellar nebula. This assumption is subject to debate, as the timescale for formation of planetesimals and embryos %(< 1.10$^6$ yr for the Joviancircumplanetary disc,  \cite{Mousis2006}) 
			can be similar to the kinetic rates of some chemical reactions \citep[$>$10$^5$ yr, see][]{Apai2010}. However, \cite{Apai2010} showed that the inefficiency of gas-grain reactions causes hot inner mid-plane regions ($<$10 AU) of the disc to reach equilibrium on a timescale of $\sim $100 years %(primarily due to endothermic reactions)
			, because chemical reactions in the gas-phase are much slower than the timescale for formation of planetesimals. The question of equilibrium is still open for the outer part of the disc ($>$10 AU), but evidence from our Solar System suggests that the assumption is valid \citep[e.g.,][]{Ebel2006, Davis2006}. 	\\
			
			%In order to compute the condensation sequence of the elements, we used the commercial software package HSC Chemistry (v. 7.1). 
			The software HSC Chemistry (v. 7.1) computes the condensation sequence of the elements under equilibrium conditions using a Gibbs energy minimization routine, thus requiring some assumptions to be made. The Gibbs energy minimum is a requirement for equilibrium in either isobaric or isothermal systems. In the presented calculations, it was assumed for simplicity that the pressure at distance \textit{r} to the star for one disc is equal to the mean value of the pressure at distance \textit{r} to the star of all the simulated discs. All simulated discs thus have the same pressure profile. Tests have shown that this is a good approximation, as discussed in Sect. \ref{disc_ab_cn}.
			Since all refractory elements have been condensed by the time the system has cooled below 200 K, the amount of refractory species at a lower temperature is thus given by their abundance at 200 K. For temperatures below 200 K, we consider only the condensation of volatile species.
						
			The refractory mineral species considered in the models are listed in Table \ref{table-ref}. It is assumed that they are formed in a solar nebula starting from neutral atoms in the gas phase and the abundances are those as given by \cite{Lodders2003} (Table \ref{table-abond}).

			\begin{table}[ht]
				\centering
				%RANGER PAR Tcond. PB: depend de la pression !
				\caption{\label{table-ref} List of Inorganic refractory phases considered in the calculations \citep{Bond2010}}
				\begin{tabular}{lll}
					\hline
					Al$_{2}$O$_{3}$ & FeSiO$_{3}$  & CaAl$_{2}$Si$_{2}$O$_{8}$  \\
					C & Fe$_{3}$P & TiC \\
					CaAl$_{12}$O$_{19}$ & NaAlSi$_{3}$O$_{8}$ & SiC \\
					Ti$_{2}$O$_{3}$ & Fe$_{3}$C & Cr$_{2}$FeO$_{4}$  \\
					CaMgSi$_{2}$O$_{6}$ & CaTiO$_{3}$ & Fe  \\
					Ca$_{2}$(PO$_{4}$)$_{2}$ & FeS & Fe$_{3}$O$_{4}$ \\
					TiN & Ca$_{2}$Al$_{2}$SiO$_{7}$ & Ni  \\
					AlN & MgAl$_{2}$O$_{4}$ & P  \\
					CaS & Mg$_{2}$SiO$_{4}$ & Si \\
					Cr & Fe$_{2}$SiO$_{4}$ & Mg$_{3}$Si$_{2}$O$_{5}$(OH)$_{4}$\\
					MgS & MgSiO$_{3}$ &  \\
					\hline
				\end{tabular}
			\end{table}

			\begin{table}[ht]
				\centering
				\caption{\label{table-abond} Initial elemental abundance of gaseous atoms in the nebula expressed in the astronomical log scale \citep{Lodders2003}.}
				\begin{tabular}{cc|cc}
					\hline
					Element & Abundance & Element & Abundance \\
					\hline
					H & $\equiv$12 & He & 10.984$\pm$0.02 \\
					C & 8.46$\pm$0.04 & N & 7.90$\pm$0.11 \\
					O & 8.76$\pm$0.05 & Na & 6.37$\pm$0.03 \\
					Mg & 7.62$\pm$0.02 & Al & 6.54$\pm$0.02 \\
					Si & 7.61$\pm$0.02 & P & 5.54$\pm$0.04 \\
					S & 7.26$\pm$0.04 & Ca & 6.41$\pm$0.03 \\
					Ti & 5.00$\pm$0.03 & Cr & 5.72$\pm$0.05 \\
					Fe & 7.54$\pm$0.03 & Ni & 6.29$\pm$0.03 \\
					\hline
				\end{tabular}
			\end{table}
		
			\subsubsection{Organic compounds}
			
			Little is known about organic refractory compounds during early solar nebula evolution. Their formation reactions and their abundances are a topic of active research. Available data are related to the %supposed
			 fractions of O, C, N, and S of the solar nebula used to form these refractory compounds \citep{Cottin1999, Lodders2003}. Direct knowledge of the chemical compositions and abundances of refractory organic materials comes mainly from studies of comets and asteroids \citep{Cottin2008}. \\
			
			Observation and detection of refractory organic molecules in the solar system are limited. The only attempts to identify these compounds are linked to the Giotto, Vega, and Stardust space missions, which analyzed ejected dust from the nucleus of the two comets  81P/Wild and 1P/Halley. For the first two missions, results indicated that instruments had a mass resolution that was too low for the identification of refractory organic molecules \citep[see][]{Huebner1987, Krueger1991, Cottin2001, Kissel2004}and data from the Stardust mission that were subject to debate due to possible contamination and warming of the dust \citep[see][]{Sandford2006, Elsila2009, Sandford2010}.
			Infrared observations have also been conducted for several comets\citep[see][]{Bockelee-Morvan1995}, and they suggest that one of the observed bands (at 3.4 $\mu$m) is linked to an organic solid component of the comets, whose chemical composition remains unknown today.\\
			
			As data are scarce for both direct and indirect identification of organic compounds, experiments were run by several research teams trying to reproduce the production of refractory organic compounds by irradiating or warming interstellar ices or cometary analogs. The rationale for such experiments is as follows: from observations of the gaseous part of the comet, one can infer the probable composition of the nucleus. To form the analogs, a mixture of key species in the gas phase is condensed onto a substrate under near vacuum conditions and then irradiated with UV photons or cosmic rays; conditions  that pre-cometary ices are assumed to have encountered in the interstellar medium. This radiation heats up the sample, and the volatile components are evaporated leaving behind the refractory organic material that can be analyzed.
			However, the refractory organic compounds detected in these experiments depend on the initial composition of the analog, the type of irradiation (UV, proton, etc...), and the analytical method used to analyze the analogs. This makes it difficult to characterize organic compounds that may form in the solar nebula. Results show the presence of polycyclic aromatic hydrocarbons (PAHs), fullerenes, carbon-chains, diamonds, amorphous carbon (hydrogenated and bare), and complex kerogentype aromatic networks. \citep[see][]{Schutte1993, Bernstein2003, MuozCaro2003, Colangeli2004, MuozCaro2004, Crovisier2005, MunozCaro2008, MunozCaro2009, Roy2010}.  \\
			
			\subsubsection{Model for formation of refractory compounds}
			
			As the presence of organic compounds in the disc and their speciation are currently unknown, two models were considered in the calculations.
	
			The first model assumes that refractory organic compounds were formed before the condensation of the nebula and do not react with other species (minerals and volatile compounds). The O, C, N, and S material used to form them is thus removed from the system at the beginning and later re-added once the condensation process is finished.
			
			The second model considers that these organic compounds do not form before the condensation commences nor during the planetary formation process. Thus, only the refractory non-organic and volatile compounds are formed during the cooling of the disc.  Atoms exist only in the form of volatile and refractory elements. 
						
			Consequently, the refractory inorganic material stays unchanged in both models, as stated in Table \ref{proportion}, which summarizes the amount of atoms that form volatile, refractory inorganic, and refractory organic components relative to their total abundance in solar nebula. Hereafter, the first model is defined as model "with" and the second as model "without".
%The second one considers that there is no formation of refractory organic molecules before and during condensation, and thus their fractions is added to the fractions of O, C, N and S used to form volatile compounds. Hereafter are defined the model "with" for calculations involving organic material formed before the condensation sequence and the model "without" for the one where organic compounds are created after the condensation sequence.
			%Table \ref{proportion} summarizes the amount of atoms for each phase and model. 
			The latter model is consistent with previous studies on this topic by \cite{Bond2010, Bond2010a}, and \cite{Madhusudhan2012a}, who do not consider the possible presence of refractory organic compounds. \\
						 
			 \begin{table}[ht]
			 	\centering
				\caption{\label{proportion}Fraction of atoms in volatile, refractory inorganic, and refractory organic components expressed as percentages of the solar nebula values. }
				\begin{tabular}{|c|c|c|c|c|c|c|}
					\hline
					 & \multicolumn{2}{|c|}{Volatile} & \multicolumn{2}{|c|}{Refractory} & \multicolumn{2}{|c|}{Refractory} \\
					 & \multicolumn{2}{|c|}{ } & \multicolumn{2}{|c|}{Inorganic} & \multicolumn{2}{|c|}{Organic} \\

					\hline
					Model & with & without & with & without & with & without \\
					\hline
					O & 59 & 74 & 26 & 26 & 15 & 0 \\
					C & 40 & 100 & 0 & 0 & 60 & 0 \\
					N & 74 & 100 & 0 & 0 & 26 & 0 \\
					S & 50 & 50 & 50 & 50 & 0 & 0 \\
					\hline
				\end{tabular}
			\end{table}
			
		\subsection{Volatile components}
			The calculation of the abundances of the volatile components is discussed in the companion paper by \cite{Marboeuf2013}. To be able to combine the refractory with the volatile content, abundances of species are normalized to the abundance of molecular hydrogen H$_{2}$ in the disc.
			The volatile components are computed by taking into account the processes of pure condensation and/or gas trapping (under the form of clathrates\footnote{Clathrates are cages of water that can trap other volatile compounds such as CO, H$_2$S and CH$_4$.}) at the surface of grains during the cooling of the disc. Calculations were run using all the processes of ice formation. Note, however, that the chemical composition of refractory components is not altered by the composition of ices. Thus, we only consider in this study the case in which gases condense as pure ices at the surface of grains. The changes induced on the chemical composition of planets by the different processes of ice formation are discussed in \cite{Marboeuf2013}.		
			%There are however small changes to the calculations on the refractory species to run more accurate calculations. These changes only interfere in the total amount of ices produced. The other results presented in the companion paper do not depend on them (position of the icelines,...).
			
		\subsection{Planetary composition}
			The first challenge is to compute the chemical composition of the planetesimals formed in the disc during cooling of the solar nebula and before the formation of planets. For this to be done,  the surface density of the disc at time t=0 (in g.cm$^{-2}$) is used, which  follows the relationship \citep{Alibert2013}:
			
			\begin{eqnarray}
				\Sigma=\Sigma_{0}.(\frac{r}{a_{0}})^{-\gamma}.e^{(\frac{r}{a_{core}})^{2-\gamma}},
			\end{eqnarray}

where $a_{0}$ is equal to 5.2 Astronomical Units [AU], $a_{core}$ is the characteristic scaling radius (in AU), $\gamma$ is the power index, $r$ is the distance to the star (in AU), and $\Sigma_{0} = (2-\gamma) \frac{M_{disc}}{2 \pi a_{core}^{2-\gamma} a_{0}^{\gamma} }$  (in g.cm$^{-2}$). The values for $a_{core}$, $\gamma$, and $\Sigma_{0}$ are varied for every disk. These characteristics of disc models are presented in Table \ref{charadisk}.
		 
		 \begin{table}[ht]
				\centering
				\caption{\label{charadisk} Characteristics of disc models \citep[from][]{Alibert2013}.}
				\begin{tabular}{|c|c|c|c|}
					\hline
					Disc & M$_{disk}$ (M$_{\odot})$ & a$_{core}$ (AU) & $\gamma$ \\
					\hline
					1 & 0.029 & 46 & 0.9 \\
					2 & 0.117 & 127 & 0.9 \\
					3 & 0.143 & 198 & 0.7 \\
					4 & 0.028 & 126 & 0.4 \\
					5 & 0.136 & 80 & 0.9 \\
					6 & 0.077 & 153 & 1.0 \\
					7 & 0.029 & 33 & 0.8 \\
					8 & 0.004 & 20 & 0.8 \\
					9 & 0.012 & 26 & 1.0 \\
					10 & 0.007 & 26 & 1.1 \\
					11 & 0.007 & 38 & 1.1 \\
					12 & 0.011 & 14 & 0.8 \\
					\hline
				\end{tabular}
			\end{table}

			Note that the temperature of the disc is {known} everywhere from 0.04 to 30 AU thanks to the planet formation model. Comparing the temperatures given by the planet formation model with the condensation sequences, the abundance of each molecule relative to H$_2$ at every distance to the star is obtained, assuming that the disc is chemically homogeneous at a given radius. That is, the condensation sequence terminates at a given temperature at a given radius.
			
			Using this method, a radial grid was computed. Whenever needed,  the chemical composition of planetesimals at a given radius is computed by interpolation on the grid by following the equation:
			
			\begin{eqnarray}
				\chi(r,i)= |1-\frac{r-r_{1}}{r_{2}-r_{1}}|.\chi(r_{1},i)+|1+\frac{r-r_{2}}{r_{2}-r_{1}}|.\chi(r_{2},i),
			\end{eqnarray}
			
where $\chi(r,i)$ is the abundance (mass fraction) at radius \textit{r} of the compound \textit{i} with $r_{1}<r_{2}$, $\chi(r_{2},i)_{r_{2}> 31 AU}\ =0$ and $\chi(r_{1},i)_{r_{1}< 0.04 AU}\ =0$. \\

			The planetary embryos placed at distance \textit{r} to the star in each of the discs have initially the same composition as the planetesimals at this distance.
		Using the planet formation model, the mass of solids accreted by the planet and the planet positions for each time step are also computed. Between two time steps $t_1$ and $t_2$, the planet has accreted a mass $\Delta m$ of planetesimals. Too get this mass transposed into composition, we use the grid so that
		
			\begin{eqnarray}
				m_{i}(t_{2}) = m_{i}(t_{1})+\Delta m.\chi(r_{t_{1}},i),   
			\end{eqnarray}
			
		where i designs compounds \textit{i}, $m_{i}(t_{1})$ is the mass of molecule \textit{i} at time step $t_1$, and $r_{t_1}$ is the distance between the star and the planet at time step $t_1$.		
		 Collisions between two planets are also considered. When they happen, planetary masses are merged without loss of solid material. Contrary to previous studies that computed the chemical composition of planets, this combination with the planet formation model makes the calculations self-consistent with the planet formation process. \\
			
	\section{Results} \label{results}
		\subsection{Ices, rocks, and iceline position}
\begin{figure} \centering
					\includegraphics[width=0.75\columnwidth]{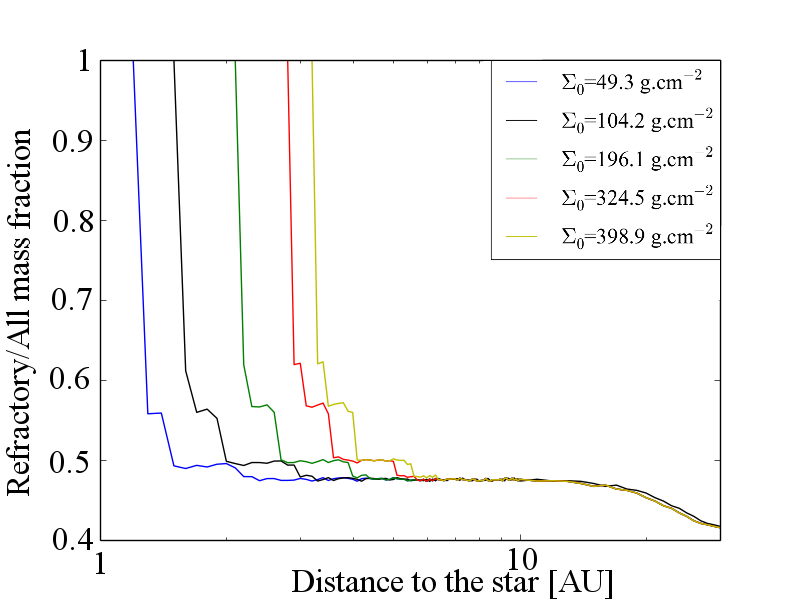}
					\caption{Refractory mass fraction relative to all condensed molecules as a function of the distance to the star in five discs of different masses (a$_{core}$ = 14 AU and $\gamma$ = 0.8, fixed). The position where the fraction drops is the position of the iceline.}
					\label{pos_iceline}
				\end{figure}

				All chemical species are different entities that condense at different temperatures. This difference creates a physico-chemical dichotomy in the disc for volatile molecules (called the "iceline"). The abundance of species thus varies depending on the distance to the star at which they condense. The iceline is then the interface position of the regions where the volatile molecule exists either in the solid or the gas phase in the disc. Between the star and the iceline, the volatile molecule exists only in the gas phase. Beyond that, the volatile molecule is in the solid state (ice).  Each volatile species has a corresponding iceline, but we call "iceline" the water iceline in this work .  Typically, the iceline for water is located between 1 and 5 AU, depending on the considered system in our calculations (see \cite{Marboeuf2013} for more details). The position of the iceline is strongly dependent on the simulated disc. The higher the mass of the disc, the higher the temperatures, and the position of the iceline is thus shifted outwards (see Figure \ref{pos_iceline}). Therefore, the composition of planets formed in these different discs changes and contributes to the diversity of planets formed. \\
				
\begin{figure} \centering
		\includegraphics[width=0.75\columnwidth]{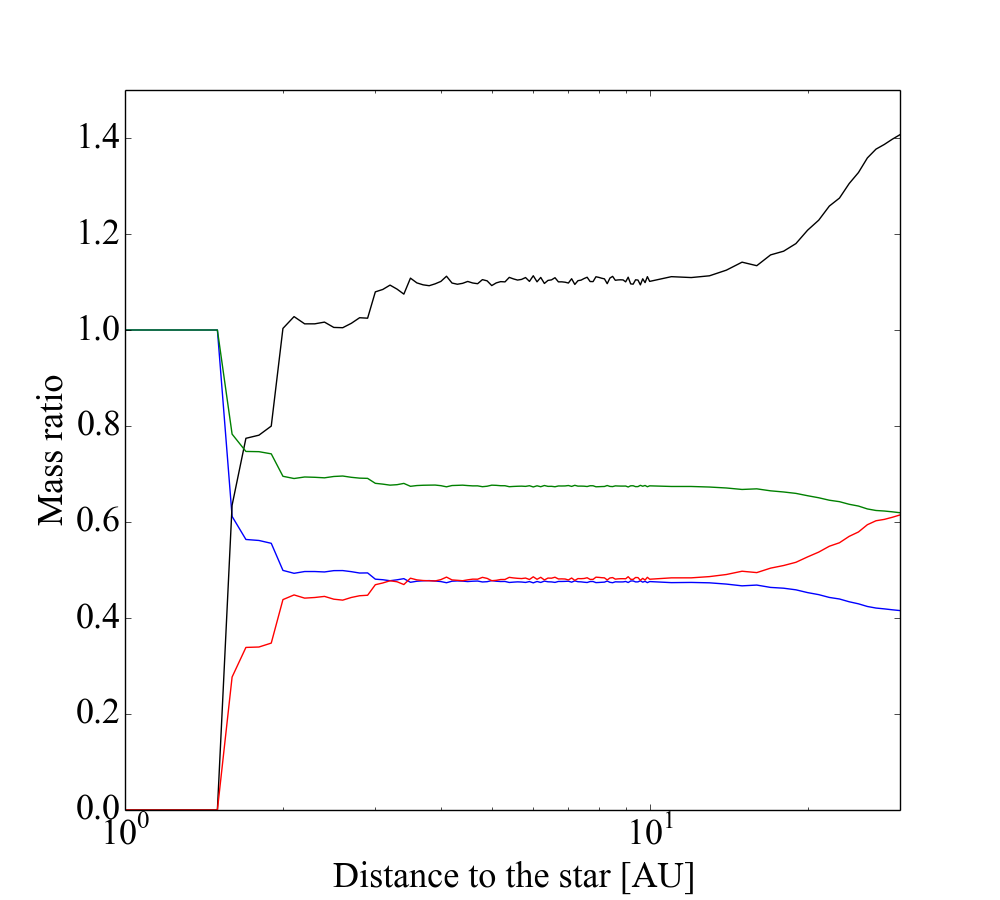} \\
		\caption{Mass fraction of refractory compounds (rocks)  relative to all condensed compounds (organic+inorganic) and ice to rock ratio as a function of distance to the star. This plot is shown for one of the simulated discs with $\Sigma_{0}$=95.844 g.cm$^{-2}$, a$_{core}$=46 AU and $\gamma$=0.9. In blue the rocks mass fraction in the model "without" is plotted; in black, we plot the ices to rocks ratio in the model "without". In green, we show the same as blue but for model "with"; in red, the same as black but for model "with"} 
		\label{refracto_vs_M_vs_A}
	\end{figure}

			Figure \ref{refracto_vs_M_vs_A} shows the ice to rock mass ratio as a function of the distance to the star and as a function of the proportion of refractory molecules relative to all molecules (refractory organic + inorganic + volatile compounds) for one of the simulated discs. The ice to rock ratio reaches values of 0.5 for the model "with" and 1 for the model "without". The model "without" is in good agreement with the  values given in \cite{McDonnell1987} for the comet 1P/Halley (mass ratio of 1) and in Greenberg's (1982) interstellar dust model \citep{Tancredi1994}. The value for the model "with" is lower and reaches the lower limit of what is observed in comets. However, these values are lower by a factor of 3 than the value generally used in models of planet formation based on \cite{Hayashi1981}.
		
		\subsection{Planet composition} \label{planet_compo}
		
\begin{figure} \centering
		\includegraphics[width=0.75\columnwidth]{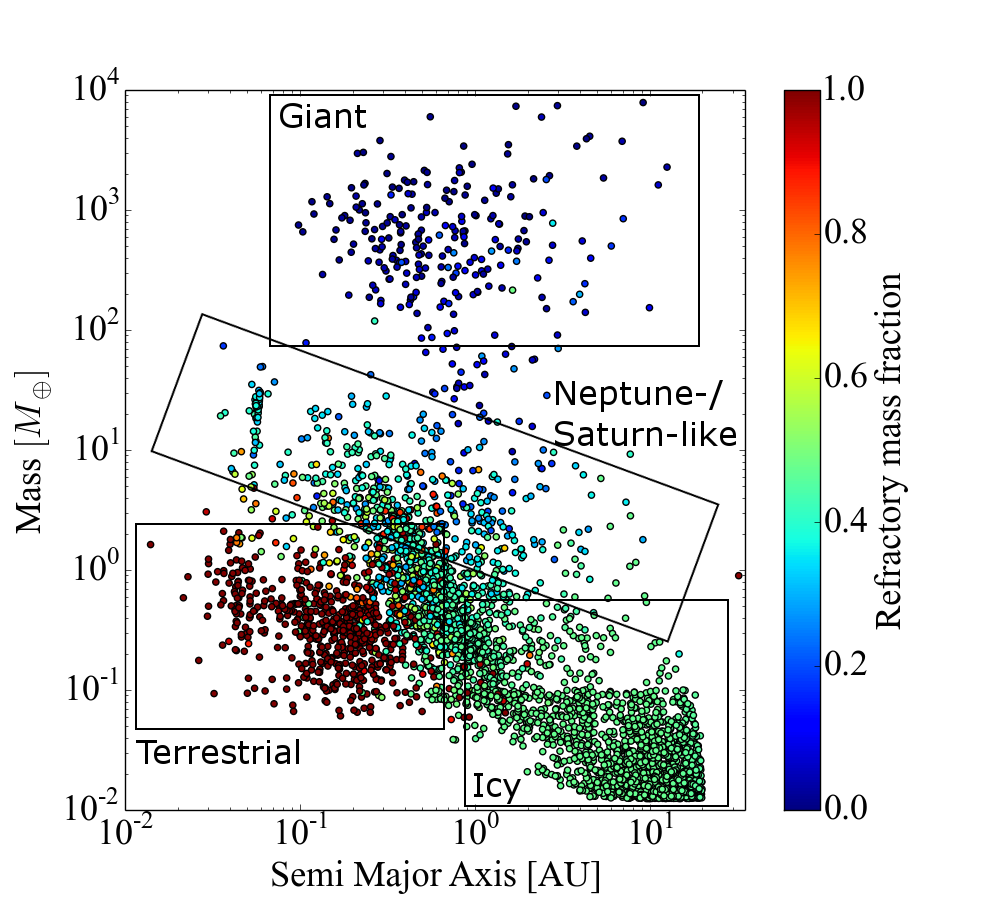}
		\caption{Refractory fraction relative to the total mass of the planet (solids + gas) for the model "without" organic compounds.}
		\label{population}
		\end{figure}

			As shown in Figure \ref{population}, the formation of different types of planets can be modeled. One group of planets with masses higher than 80 M$_ {\oplus}$ consists of giant planets with a large envelope of gas. Their masses are dominated by gas and the refractory mass fraction is thus close to 0 wt \%. The second population  contains up to 60 wt \% of refractory elements. These planets, whose masses are between 10 and 100 M$_{\oplus}$, are likely to be Neptune- or Saturn-type planets. The last two populations have masses similar to that of the Earth or smaller. They are icy planets or ocean planets, and they can possibly have water at their surface. The rocky planets have low masses  and include planets from 0.01 M$_{\oplus}$ up to 4 M$_{\oplus}$ with a semi-major axis smaller than 1 AU.  They represent 19\% of the simulated planets. The icy, Neptune-like, Saturn-like, and giant gas planets are considered as one population (population B) for the rest of this study.  This contribution focuses mainly on the rocky population, hereafter population A, as their refractory mass fraction is higher. The composition of the other planets is studied in more detail in the companion paper of \cite{Marboeuf2013}. \\

		\begin{figure} \centering
		\subfloat[Model "with"]{\includegraphics[width=0.75\columnwidth]{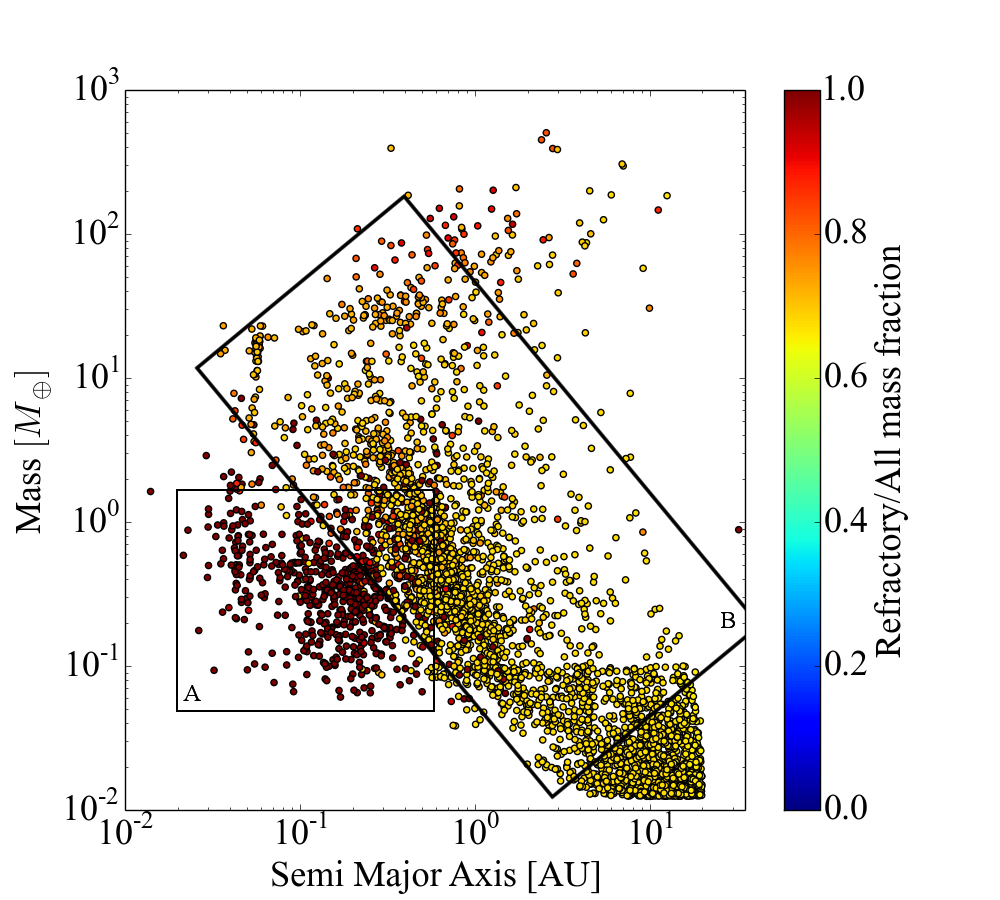}}\\
		\subfloat[Model "without"]{\includegraphics[width=0.75\columnwidth]{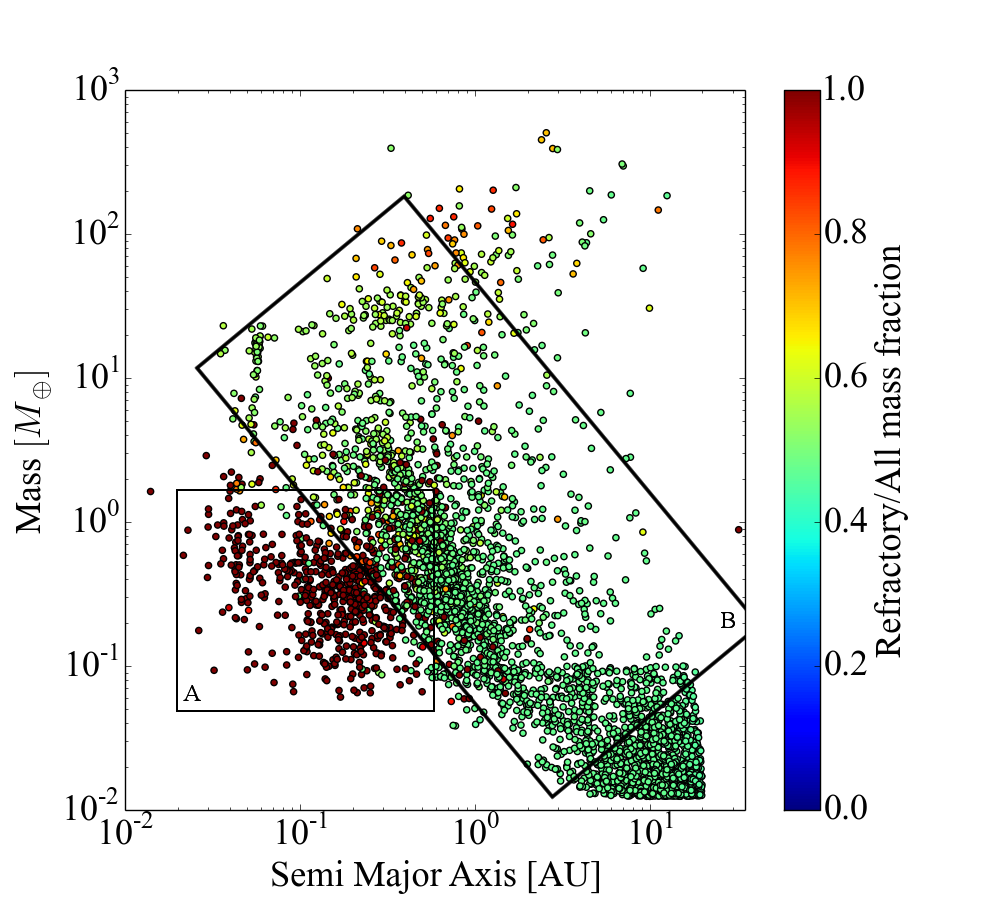}}
		\caption{Refractory mass fraction relative to all condensed molecules  as a function of the mass of solids accreted by the planet and the final position of the planets. The top panel (a) shows results for the model "with" refractory organics and the bottom panel (b) shows results for the model "without" refractory organics (see Table \ref{proportion})}  
		\label{refracto2_vs_M_vs_A}
	\end{figure}

			Figure \ref{refracto2_vs_M_vs_A} shows the mass fraction of refractory species (organic + inorganic compounds) relative to all condensed elements (volatile + organic + inorganic compounds) for each of the simulated planets. Calculations show that all the simulated planets have a refractory mass fraction greater than 50 wt \%. Planets of population A are nearly all rocky. These planets, which formed near the star, had not migrated beyond the iceline, where volatile compounds are present. All of the other simulated planets (i.e., population B) are composed of $\sim$70 wt \% of refractory elements in the solids  for the model "with" and $\sim$40 wt \% for the model "without". This difference between each case can be explained by the presence of refractory organic compounds in the model "with", which is lacking in the model "without".  The higher mass of the planets and/or their higher semi-major axis allow them to accrete more volatile compounds and gas. This growth occurs because during their formation they spend an appreciable amount of time beyond the iceline (typically 1-5 AU). %A depletion by a factor of 2 can be observed for Mg, Ca, Na, Si and Fe, as well as an enrichment in volatile elements such as C and O by a factor 4 to 10.
				
\begin{figure} \centering
		\subfloat[Model "with"]{\includegraphics[width=0.75\columnwidth]{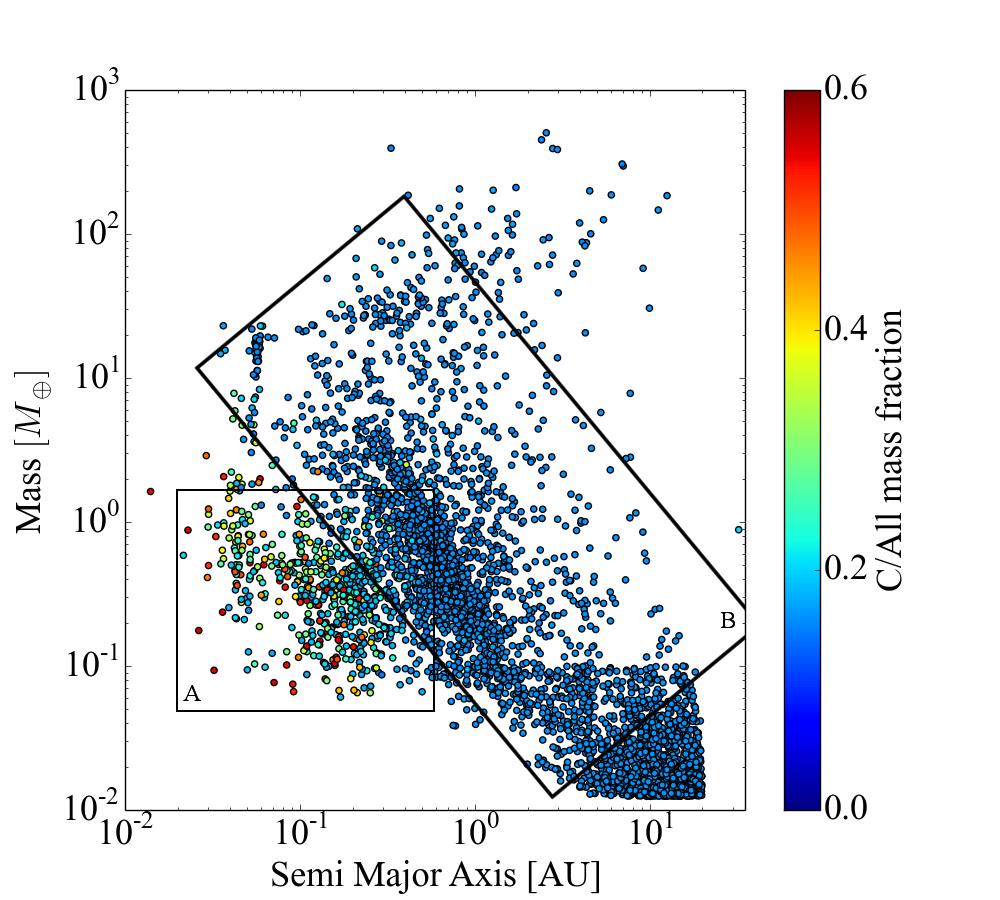}} \\
		\subfloat[Model "without"]{\includegraphics[width=0.75\columnwidth]{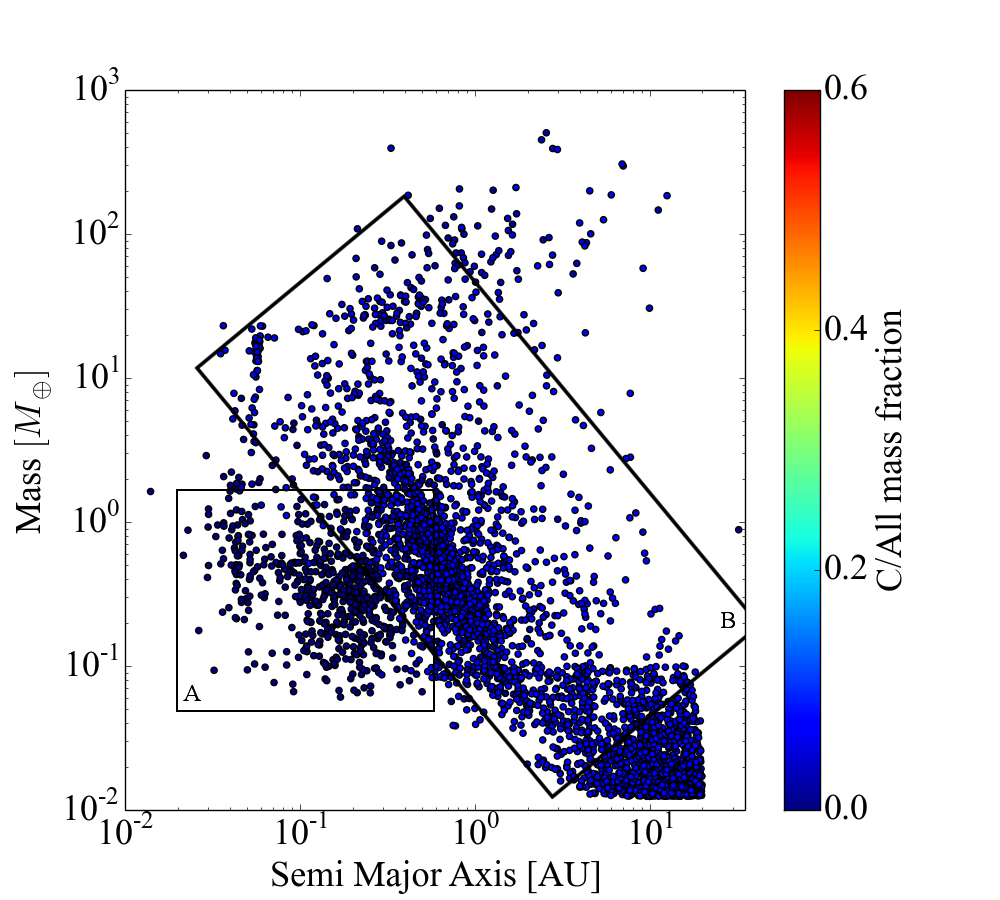}}
		\caption{Same as Figure \ref{refracto2_vs_M_vs_A} but for the C mass fraction.} 
		\label{C_vs_M_vs_A}
		\end{figure}

			Figure  \ref{C_vs_M_vs_A} shows the mass fraction of C relative to all condensed elements plotted as a function of the mass of solids and the final position of the planets. %As one can see in the top panel, for planets of population A, 
			As shown in Figure \ref{C_vs_M_vs_A}a, this fraction can vary from nearly 20\% up to 60\% for population A, while the mass fraction of C stays at an average value of 20\% in the model "with" for population B. The model "without" is more homogeneous with values ranging from less than 1\% for population A up to 6-8\% for population B (Figure \ref{C_vs_M_vs_A}b).  			
									
\begin{figure} \centering
		\subfloat[Model "with"]{\includegraphics[width=0.75\columnwidth]{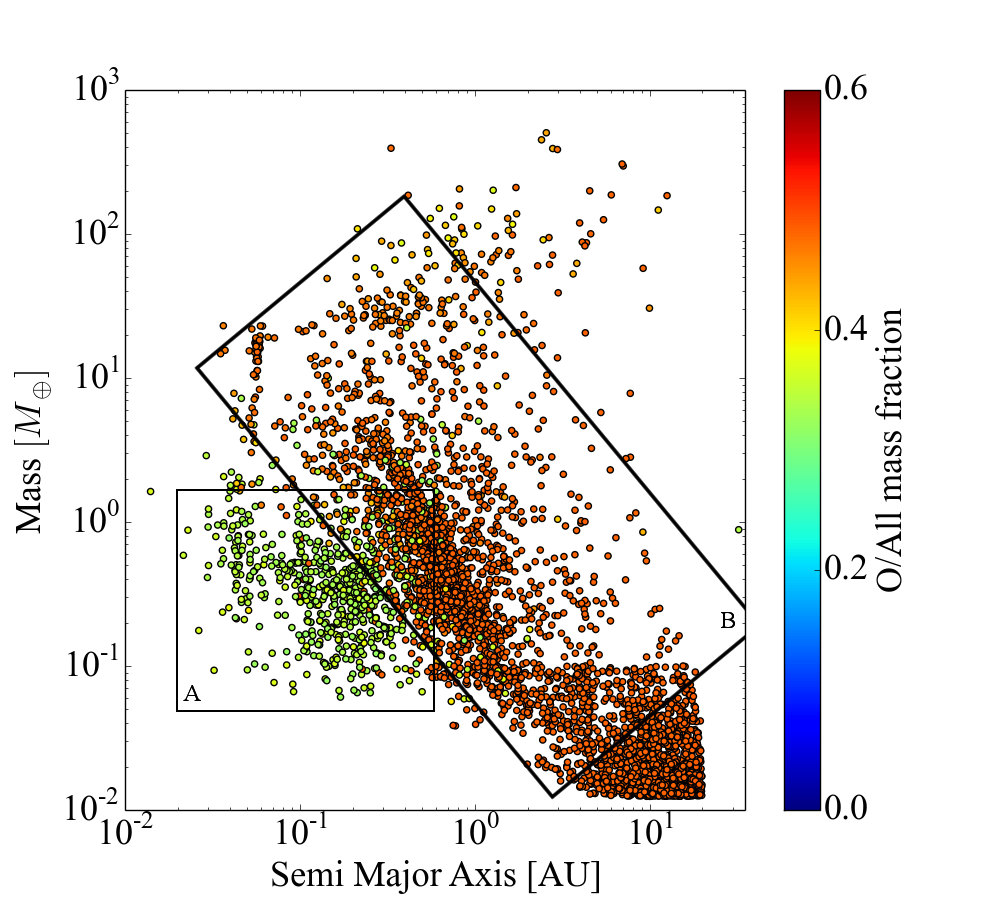}} \\
		\subfloat[Model "without"]{\includegraphics[width=0.75\columnwidth]{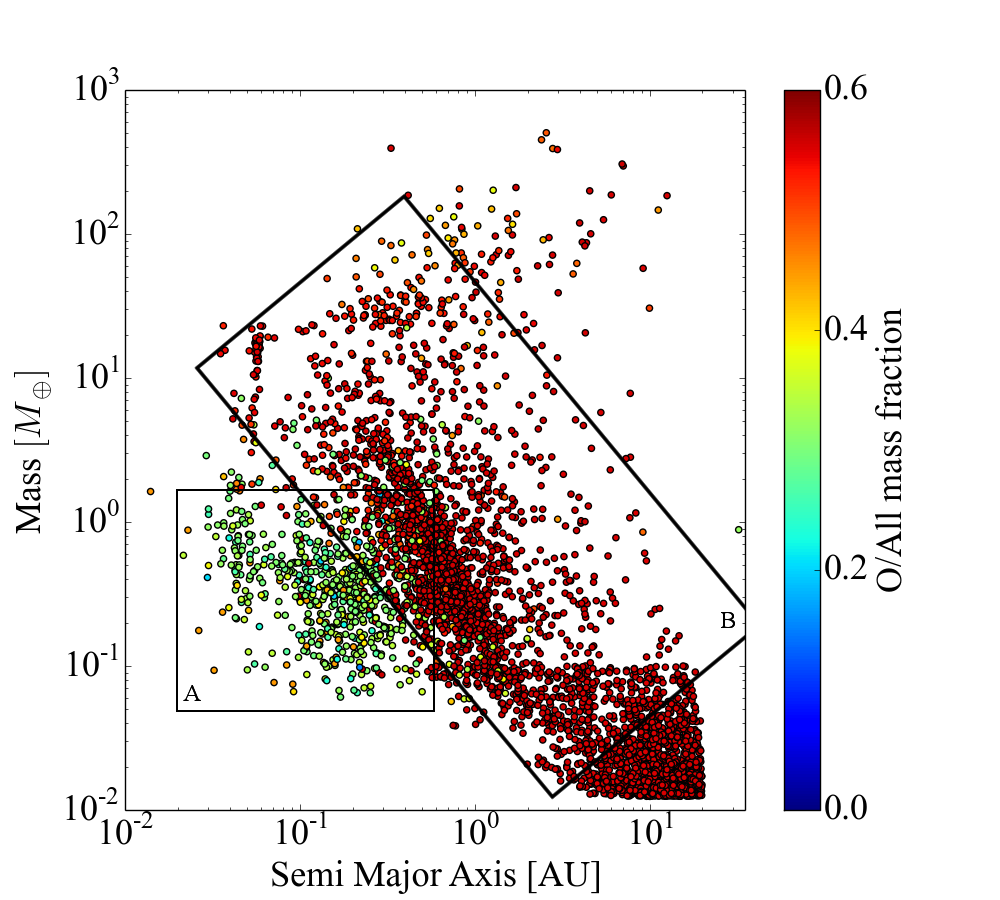}} 
		\caption{Same as Figure \ref{refracto2_vs_M_vs_A} but for the O mass fraction.} 
		\label{O_vs_M_vs_A}
	\end{figure}

			The abundance of O relative to all condensed elements is shown Figure \ref{O_vs_M_vs_A}. This abundance remains almost constant at around 30-35 wt \% of O for population A for both models and for 53 - 60 wt \% for population B. However, some planets have a higher mass fraction in population A for the model "without". This can be explained by the assumption that refractory organics conserve C, O, N, and S atoms during the evolution of the disc in the model "with", which is not the case in the model "without" due to the loss of material in volatile species (see Marboeuf et al. 2013). There are then fewer atoms of C, O, N, and S in the model "with".
			
\begin{figure} \centering
		\subfloat[Model "with"]{\includegraphics[width=0.75\columnwidth]{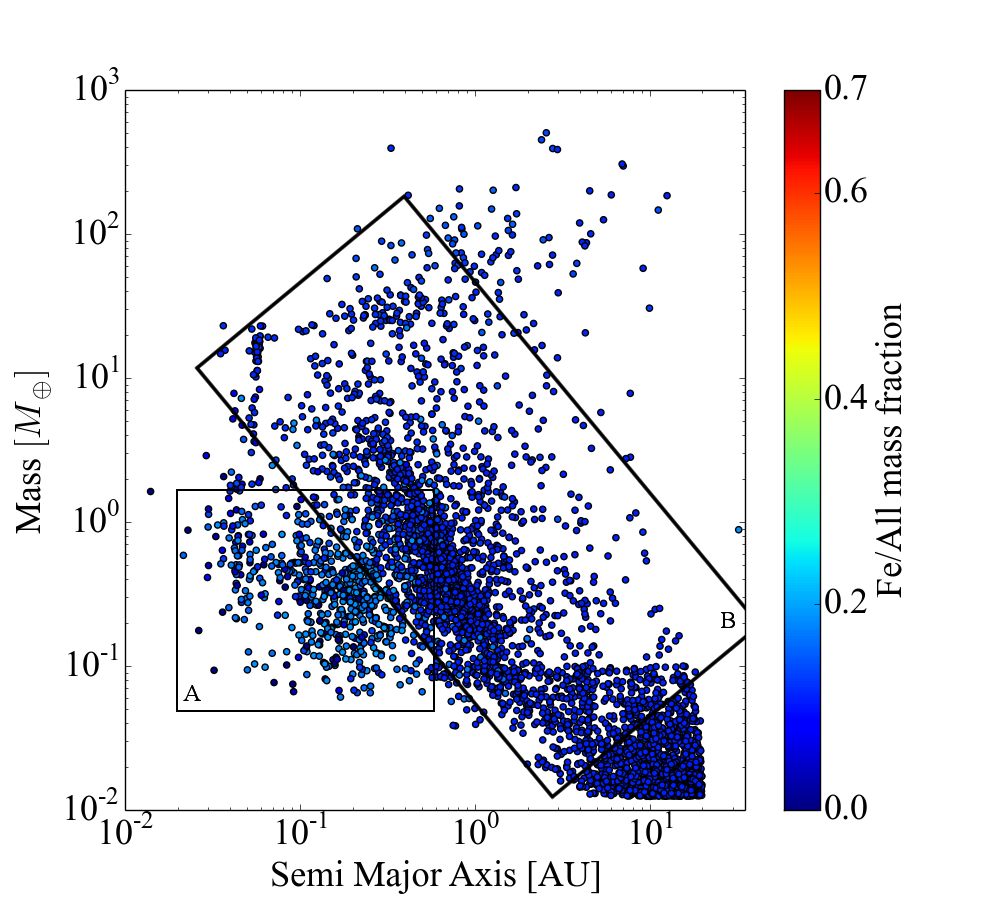}}\\
		\subfloat[Model "without"]{\includegraphics[width=0.75\columnwidth]{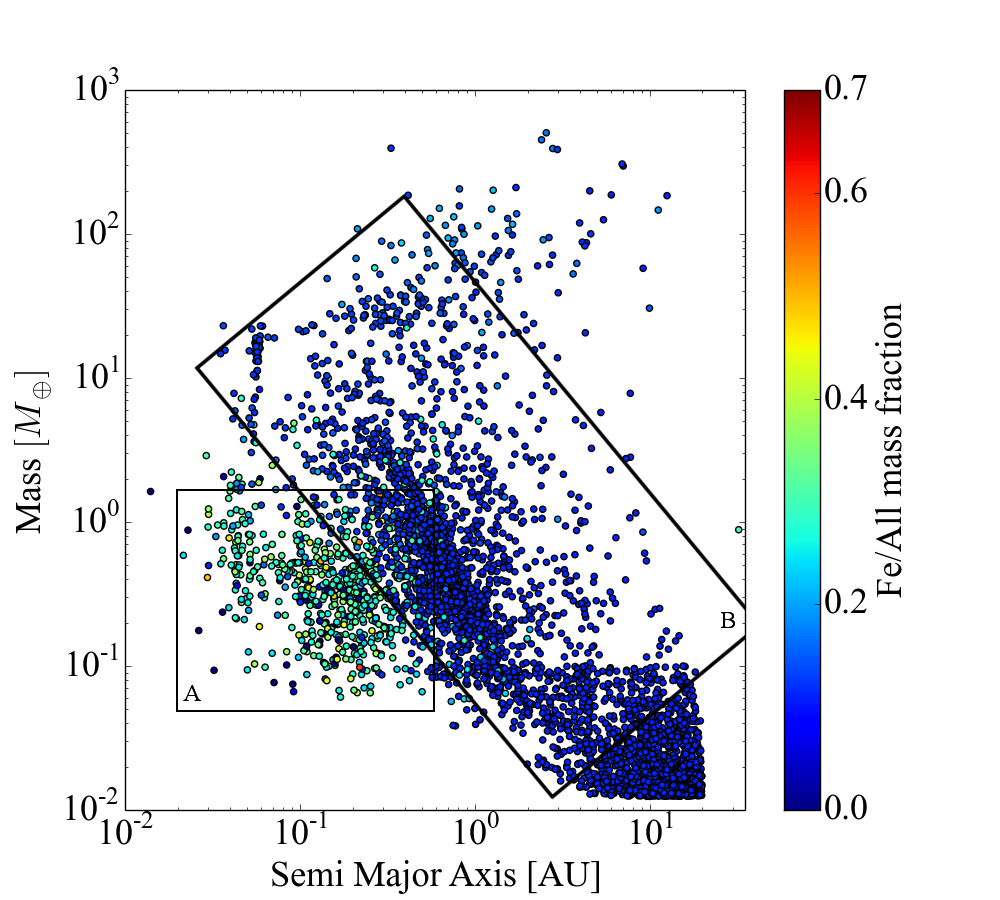}}
		\caption{Same as Figure \ref{refracto2_vs_M_vs_A} but for the Fe mass fraction.}  
		\label{Fe_vs_M_vs_A}
	\end{figure}

			As shown in Figure \ref{Fe_vs_M_vs_A}, the mass fraction of Fe ranges from 20 wt \% for model "with" to 40 wt \% for model "without" for population A and between 10 and 15 wt \% for population B. Some exceptions in population A can be found, with up to 60 wt \% of Fe.

\begin{figure} \centering
		\subfloat[Model "with"]{\includegraphics[width=0.75\columnwidth]{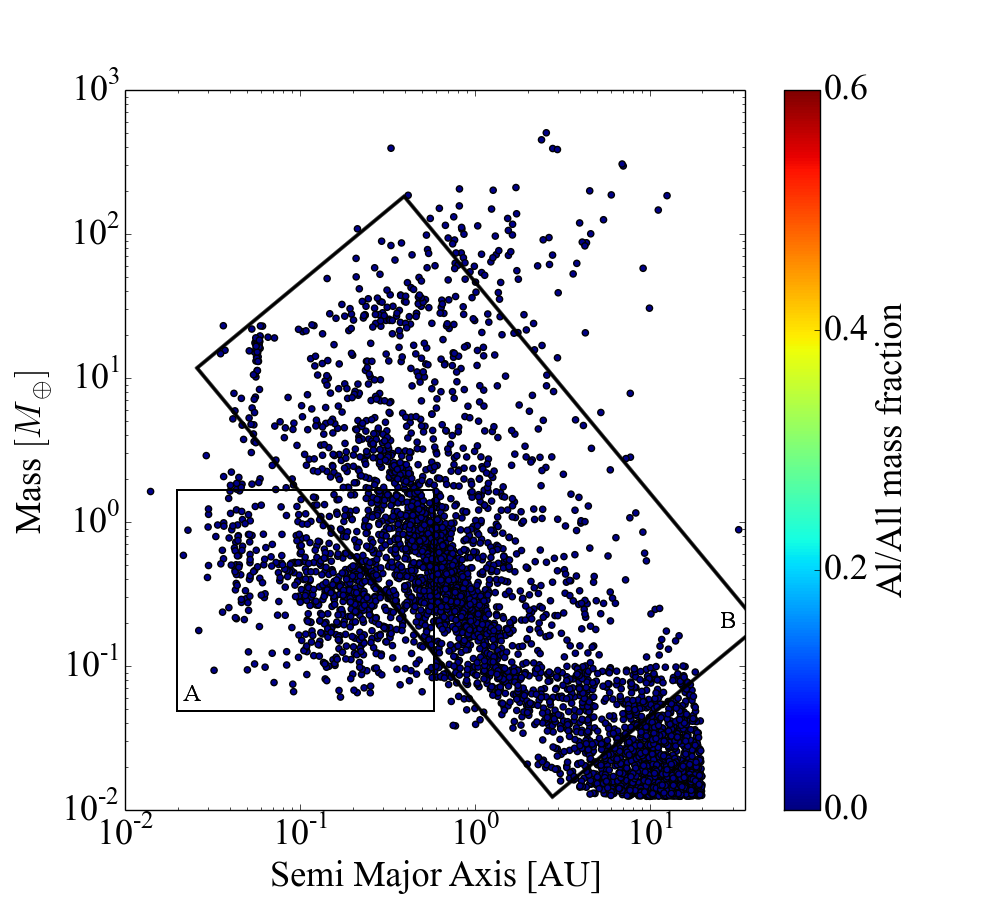}} \\
		\subfloat[Model "without"]{\includegraphics[width=0.75\columnwidth]{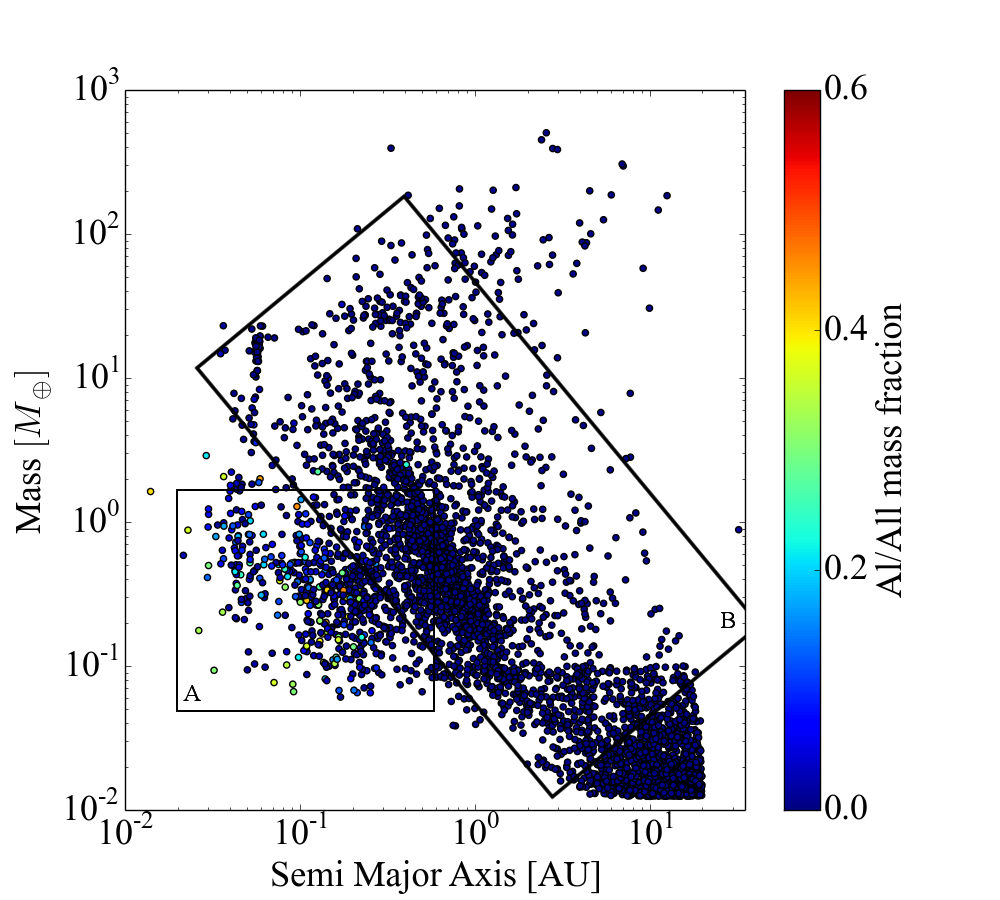} }
		\caption{Same as Figure \ref{refracto2_vs_M_vs_A} but for the Al mass fraction.} 
		\label{Al_vs_M_vs_A}
	\end{figure}

			The Al  abundance is not very high in both populations (Figure \ref{Al_vs_M_vs_A}). It rarely exceeds 2 wt \% of the mass of solids in the model "with" and stays at values close to 0 wt \% for the model "without". Some exceptions can be seen with planets in population A having 50 wt \% of Al that corresponds to planets with a mass fraction close to 0 wt \%  for Fe (Figure \ref{Fe_vs_M_vs_A}). %This indicates that if rocky planets do not contain iron, this element is replaced by aluminum mainly.  

\begin{figure} \centering
		\subfloat[Model "with"]{\includegraphics[width=0.75\columnwidth]{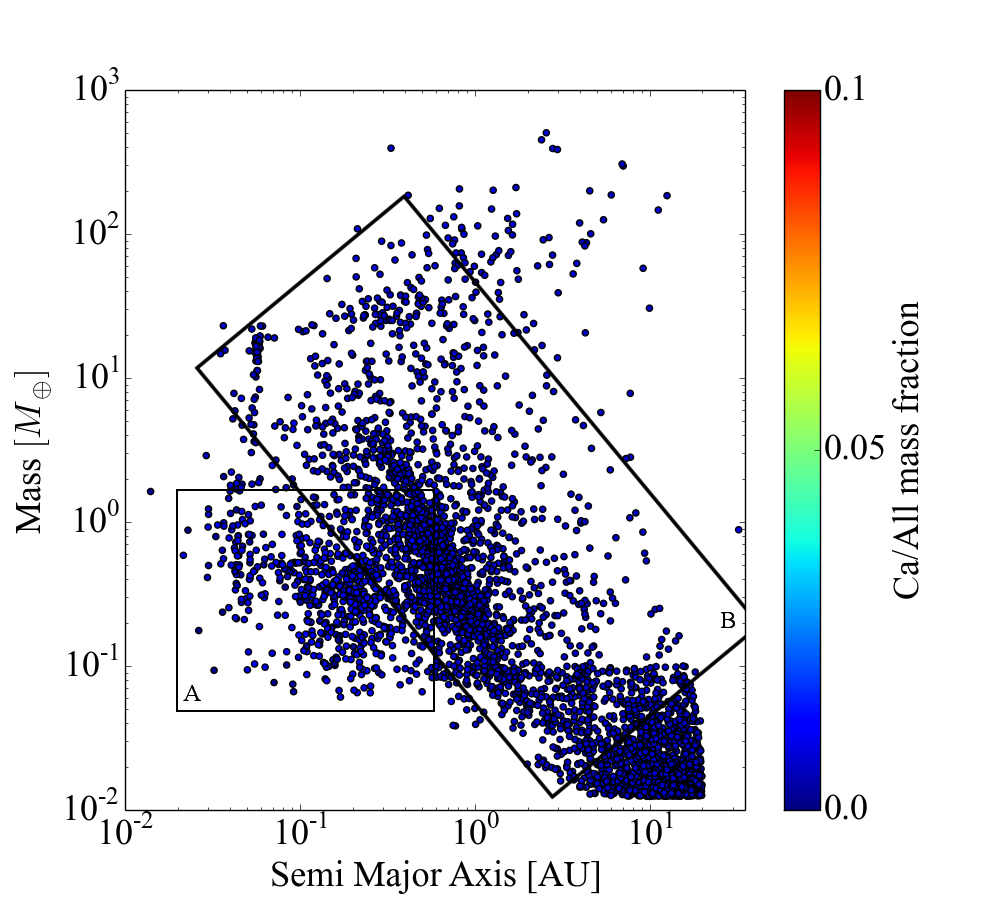}} \\
		\subfloat[Model "without"]{\includegraphics[width=0.75\columnwidth]{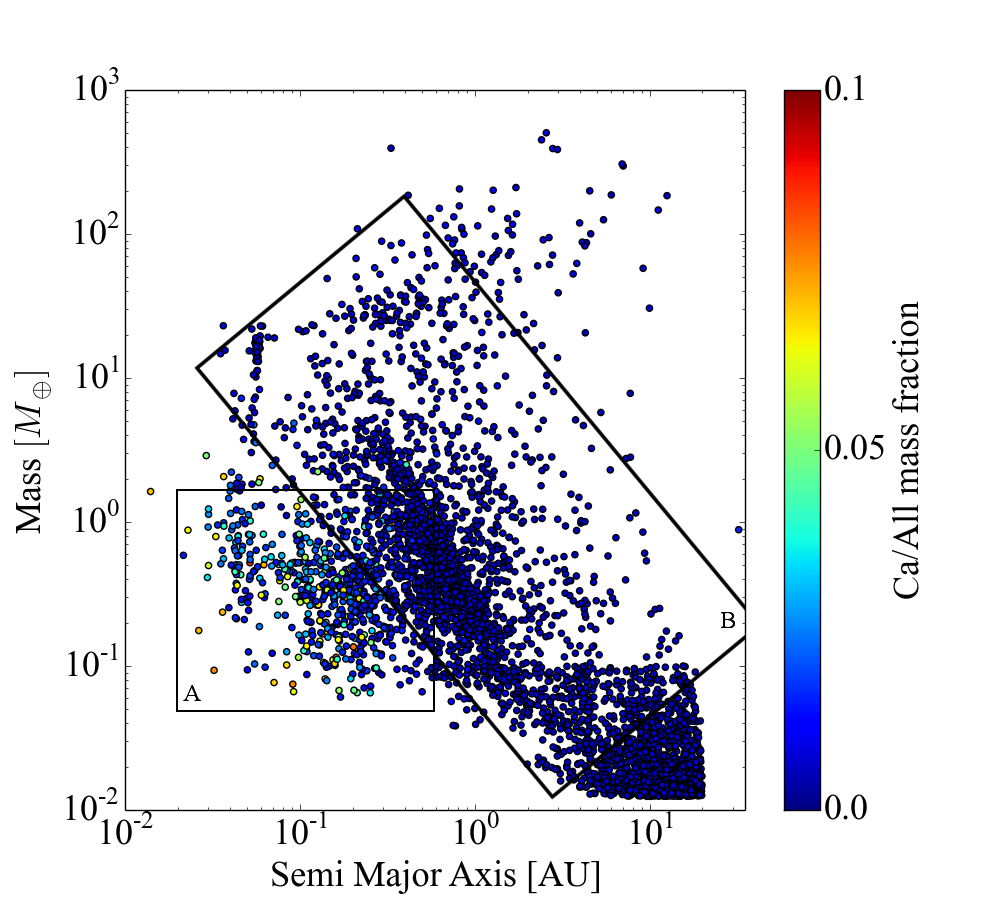}} 
		\caption{Same as Figure \ref{refracto2_vs_M_vs_A} but for the Ca mass fraction.} 
		\label{Ca_vs_M_vs_A}
	\end{figure}

\begin{figure} \centering
		\subfloat[Model "with"]{\includegraphics[width=0.75\columnwidth]{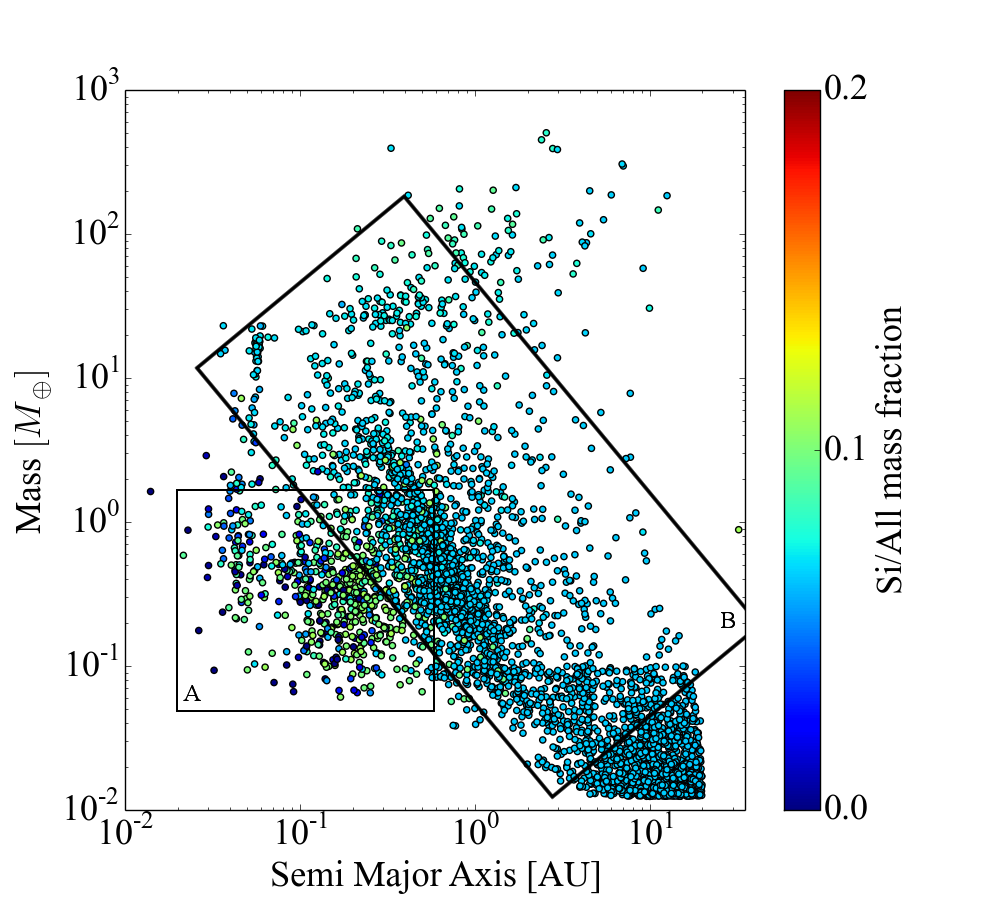} }\\
		\subfloat[Model "without"]{\includegraphics[width=0.75\columnwidth]{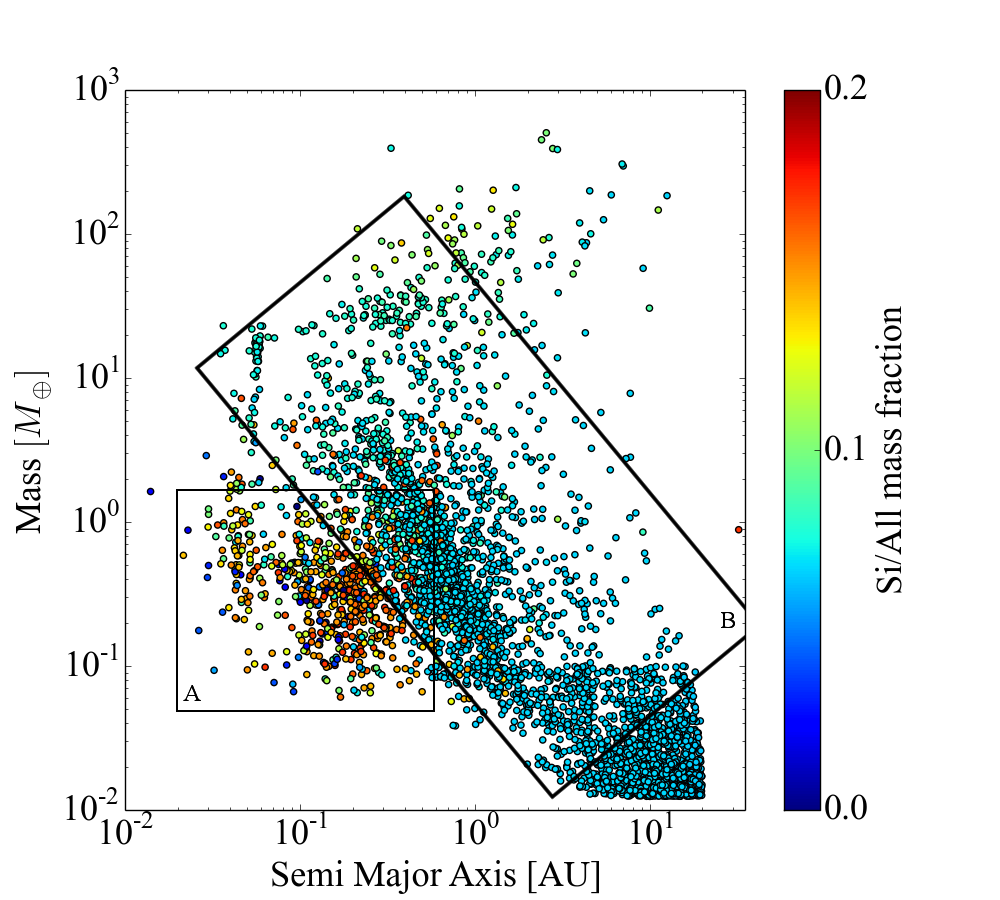} }
		\caption{Same as Figure \ref{refracto2_vs_M_vs_A} but for the Si mass fraction.} 
		\label{Si_vs_M_vs_A}
	\end{figure}

			Figures \ref{Ca_vs_M_vs_A} and \ref{Si_vs_M_vs_A} show the mass fraction of Ca relative to all condensed elements and the mass fraction of Si relative to all condensed elements respectively. The Ca composition does not change significantly between models, but some planets from population A have up to 9 wt \% of Ca. Comparing with the Al abundances, one can see that they are the same as planets with high Al mass fraction.
			
			Silicon is an important element  as it forms silicates that can be found in the Earth's crust and mantle. Figure \ref{Si_vs_M_vs_A} shows the Si mass fraction in the simulated planets. Results are identical for population B in both models; however, an increase by a factor of 2 can be found for planets of population A between model "with" and model "without".
			

\begin{figure} \centering
		\subfloat[Model "with"]{\includegraphics[width=0.75\columnwidth]{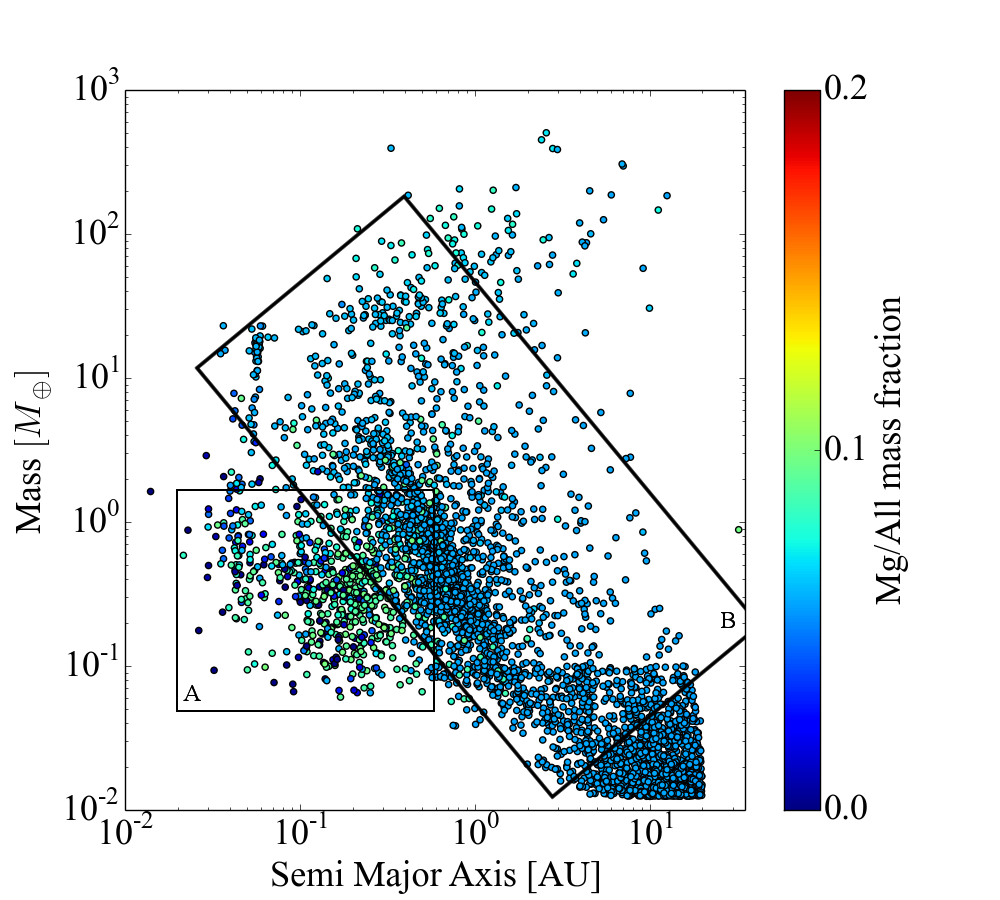}}\\
		\subfloat[Model "without"]{\includegraphics[width=0.75\columnwidth]{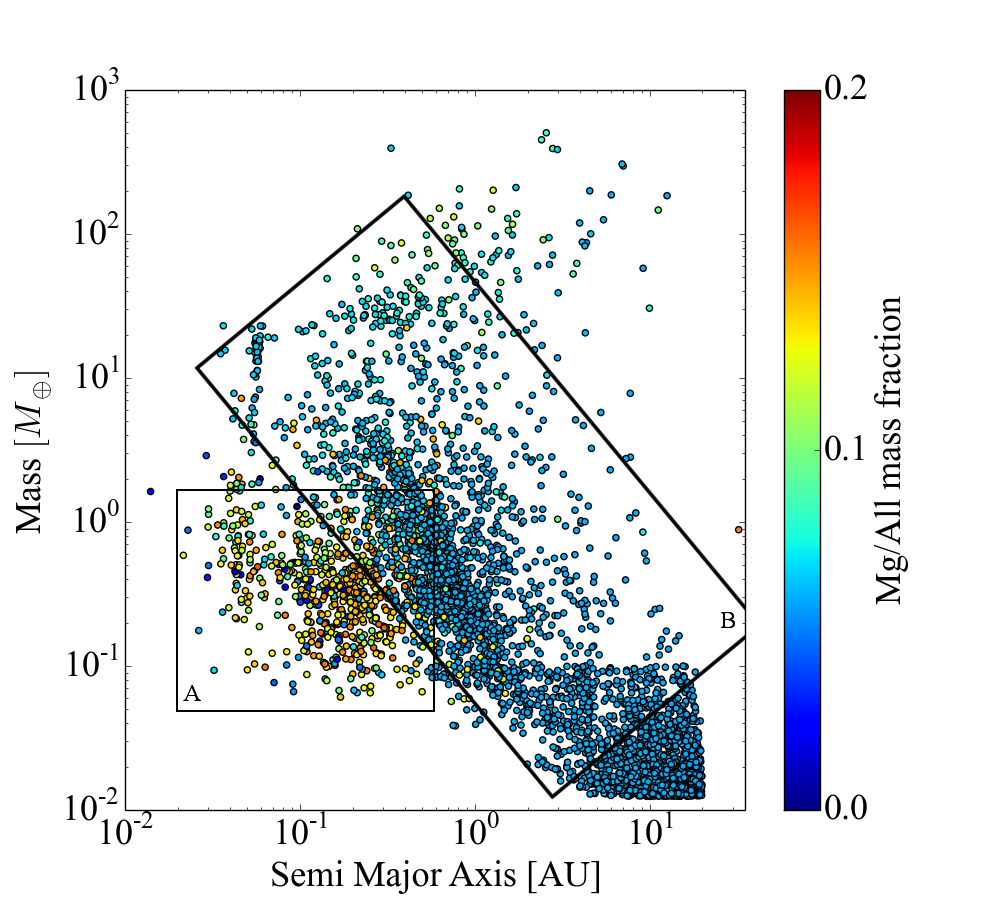}}
		\caption{Same as Figure \ref{refracto2_vs_M_vs_A} but for the Mg mass fraction.}  
		\label{Mg_vs_M_vs_A}
	\end{figure}

\begin{figure} \centering
		\subfloat[Model "with"]{\includegraphics[width=0.75\columnwidth]{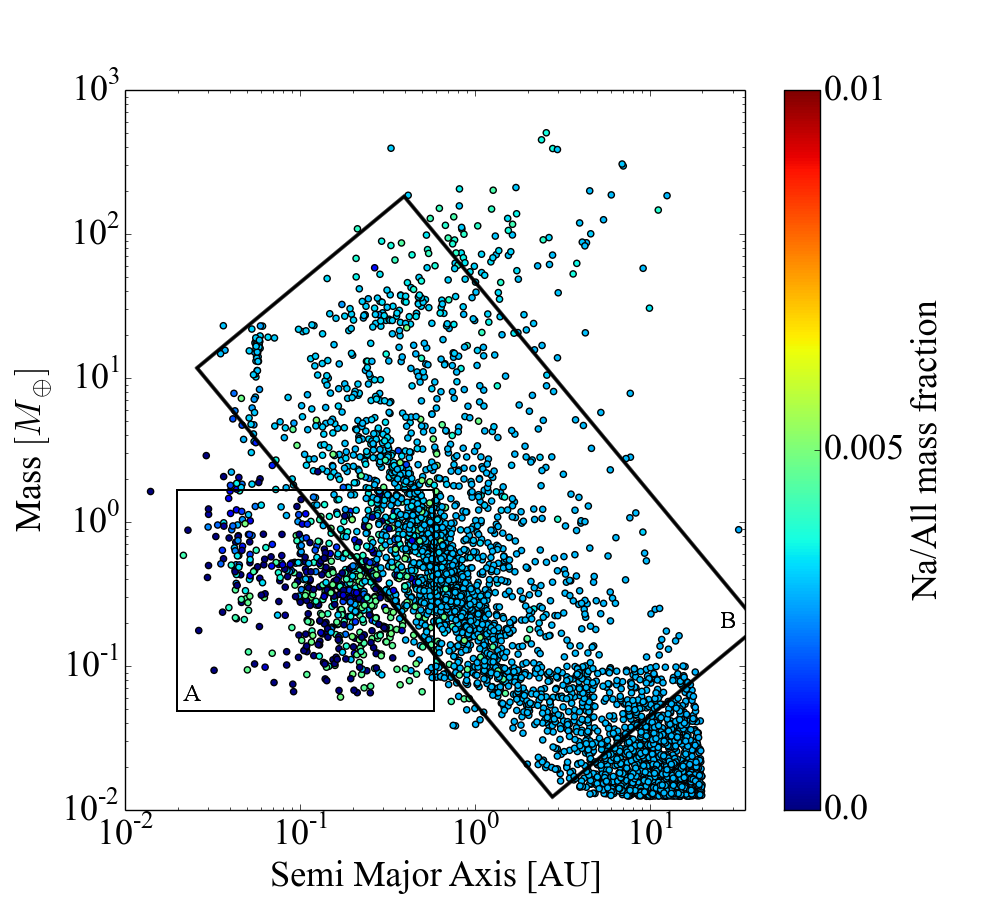}}\\
		\subfloat[Model "without"]{\includegraphics[width=0.75\columnwidth]{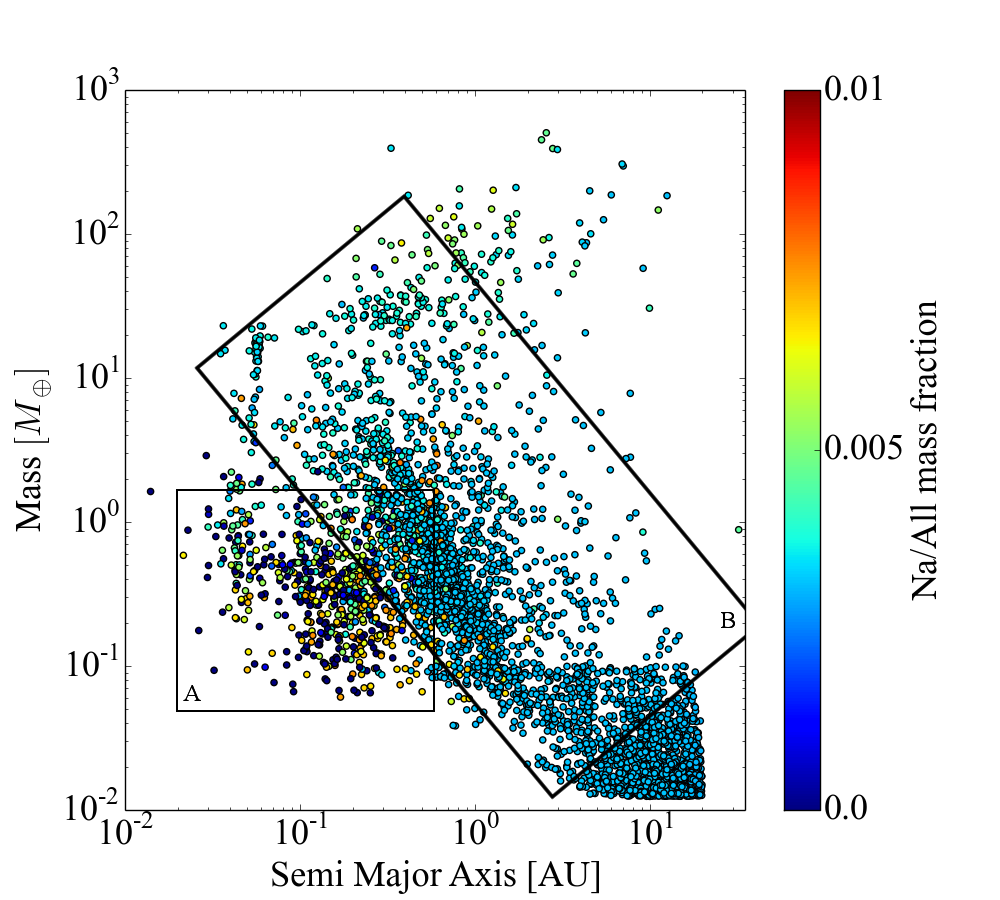} } 
		\caption{Same as Figure \ref{refracto2_vs_M_vs_A} but for the Na mass fraction.} 
		\label{Na_vs_M_vs_A}
	\end{figure}

			The last two figures represent the mass fraction of Mg (Figure \ref{Mg_vs_M_vs_A}) and that of Na (Figure \ref{Na_vs_M_vs_A}) compared to all condensed elements. The abundance of Na does not change much, but rocky planets contain slightly more (by a factor of 2) Na. The Mg mass fraction is more interesting, as Mg is incorporated in olivine and pyroxene minerals. Magnesium can be found in similar abundance to that of the Si on Earth, which is the case in the calculations, with 5 wt \% of Mg in population B and between 10 and 20 wt \% in population A. A general decrease of its mass fraction between the model "without" and the model "with" exists. \\			
	
	
		Table \ref{recap_ab} summarizes the typical and unusual abundances described with the plots.
			
		\begin{table}[ht]
			\centering
			\caption{\label{recap_ab}Typical and unusual elemental abundances for rocky planets (population A) derived in this work in wt \%. The dash indicates that there is no values to report.}
			\begin{tabular}{|c|c|c|}
			\hline
			Atom & Typical abundance & Unusual abundance \\
			\hline
			C & 0-10 & 40-60  \\
			O & 30-35 & - \\
			Fe & 30-35 & 0; 60-70 \\
			Al & 0-2 & 40-50 \\
			Ca &1-2 & 7-9 \\
			Si & 10-20 & 2-4 \\
			Mg & 10-20 & 2-4 \\
			Na & 0-0.9 & - \\
			\hline
			\end{tabular}
		\end{table}

			\paragraph{{Solar System-like planets}}
			\begin{table}[ht]
				\centering
				\caption{\label{planet_ss}Planetary abundances in wt \%(Fe, O, Mg, Al, Si, Ca) or weight ppm (C, Na) of Mercury and Venus \citep{Morgan1980}, Earth \citep{Kargel1993} and Mars \citep{Lodders1997}}
				\begin{tabular}{|c|c|c|c|c|}
					 \hline
					  & Mercury & Venus & Earth & Mars \\
					  \hline
					  Fe & 64.47 & 31.17 & 32.04 & 27.24 \\
					  O & 14.44 & 30.9 & 31.67 & 33.75 \\ 
					  Mg & 6.5 & 14.54 & 14.8 & 14.16\\
					  Al & 1.08 & 1.48 & 1.43 & 1.21\\
					  Si & 7.05 & 15.82 & 14.59 & 16.83\\
					  Ca & 1.18 & 1.61 & 1.6 & 1.33\\
					  C & 5.1 & 468 & 44 & 2960 \\
					  Na & 200 & 1390 & 2450 & 5770\\
					  \hline	
				\end{tabular}
			\end{table}
			
\begin{figure} \centering
				\includegraphics[width=0.75\columnwidth]{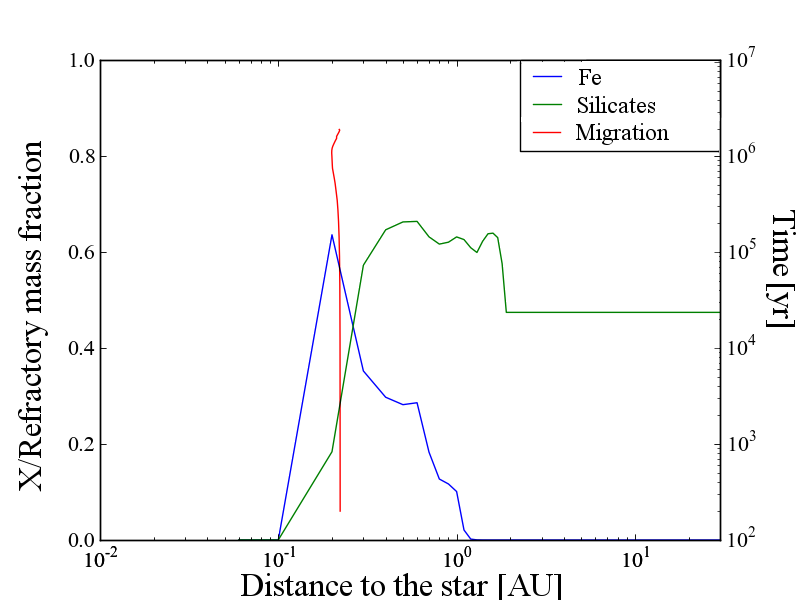} \\
				\caption{Evolution of a Mercury-like planet in its disc (a$_{core}$ = 46 AU, $\Sigma_0$ = 213.86 g.cm$^{-2}$, $\gamma$ = 0.9). The migration of the planet through its life happened between 0.22 and 0.25 AU, where atomic Fe condensed in higher quantities than silicates ($\sim$ 65 wt \% against $\sim$30 wt \%). Model "without".} 
				\label{mercury}
			\end{figure}

			Table \ref{planet_ss} shows the abundances of major elements in the terrestrial planets of the Solar System. The concentrations of the elements for Venus, Earth, and Mars are often reproduced in the calculations. Interestingly, the average composition of Mercury can also be reproduced, though such planets are rare and only 0.4\% of the planets end up with a composition similar to Mercury.	
			%The planets that match Mercury's abundances are less than 20. Mercury-like planets can be modeled as a product of condensation and there is no need to explain their relatively high Fe-content as a consequence of impact erosion (see, e.g. \cite{Benz2007}). 
			The compositions of the discs in which such planets are formed have been studied in more details, and it can be shown that such planets must have been migrating in regions where atomic Fe had condensed before silicates condensed (for an example see Figure \ref{mercury}). This is a similar result to \cite{Lewis1972}, who considered that the anomalous Fe content of Mercury was due to fractionation of Fe from silicates during the formation process itself. However, \cite{Lewis1988} later recognized that this scenario does not explain the Fe abundance of Mercury.  The dilution of high density material with low density material was not taken into account in \cite{Lewis1972}. Including this dilution effect, \cite{Lewis1988} showed that the Fe content is  reduced by half. This result highlights one of the processes not considered so far: the possible drift of planetesimals.	
			Differences in the elemental content in planets can be seen between the model "with" and the model "without". For example, the results in Figure \ref{C_vs_M_vs_A} applied to population A is closer to the C mass fraction in the Earth's solids, which is estimated to be of about 1 wt \%  \citep{Morgan1980}. {Carbon seems to be completely present only in volatile compounds along with O. In general, the model "without" better reproduces the elemental abundances of terrestrial planets in the solar system.}

			\paragraph{{Al-rich terrestrial planets}}
			As can be seen from Table \ref{recap_ab}, there are some unusual planets that can be formed amongst population A. In the model "without", planets with no Fe (or negligible Fe) are usually highly enriched in Al with a concentration of up to 50\% of the mass of solids in Al. The O plot shows that these planets are also rich in O with concentration up to 43 wt \%. Calcium is the only other significant element present with 7 wt \%. The calculations of disc properties and the evolution of these planets show that they were formed in a massive disc, hence, the temperature was higher than in other discs considered here, and the only species formed are Al, O, and Ca condensates.\\
			%Other species are 7\% of Ca and faint traces of other elements. 
			
			For example, the planet located at $\sim$ 0.03 AU with a mass of about 0.1 M$_{\oplus}$ moved from 0.15 AU at t=0 to 0.03 AU at the end of the calculations. Temperatures at these radii were $\sim$1500 K and 1700 K respectively. At 1500 K, only Al$_{2}$O$_{3}$ and Ca-Al-O bearing phases condense at the corresponding pressure of $\sim$ 10$^{-4}$ bars; thus, the solids accreted by this planet are mainly composed of Al, O, and Ca. Iron species do not condense, which explains why there is no Fe in these planets.
			
			\paragraph{{C-rich terrestrial planets}}
			 In the model "with", an enrichment in C with a concentration up to 60\% of the mass of the solids can be observed in certain planets. This enrichment can be explained by the presence of organic compounds in the model. Only Ca-Al-O complexes are able to condense at the high temperatures prevalent in the discs in which these planets form, but the fraction of C, O, N, H, and S combined into organic compounds (which are considered to form before the cooling of the nebula and added later) is much higher than the accreted mass of Ca, Al and, O. Thus, the relative amount of O has slightly decreased but remains high, while Al and Ca only represent less than 5 wt \% of the mass of solids. Carbon, however, becomes the most abundant component. Such planets are thus extremely rich in organic compounds and a question arises: is it really possible to form refractory organic compounds in such high quantity before the condensation of the stellar nebula ? 

			%\paragraph{{Conclusion}}
			Both models show that a wide variety of compositions can be formed. Starting from a similar initial nebula composition, the calculations show that planets can be compositionally very different from each other. This is particularly the case for rocky planets, whose differences are much more pronounced than for gas giants or ocean planets. Consequently, the diversity of planets formed is the result of the planet formation process itself and depends on the disc properties (i.e., mass, temperatures, density...) and on the initial abundances of elements (Figures \ref{pos_iceline} and \ref{refracto_nonself}).
			
		\subsection{Distribution of carbides} \label{sect_carb}
			One of the most important elemental ratios for the mineralogy of exoplanets is the C/O ratio. This ratio controls the distribution of carbides and silicates in the system. For ratios higher than 0.8,  Si is expected to be mainly used to form mainly solid carbide (SiC), as the system is C-rich. If the C/O ratio is lower than 0.8, Si combines with O to form silicates, which are basic building blocks of rock forming minerals, such as SiO$_{2}$ or SiO$_{4}^{4-}$. \\
			
\begin{figure} \centering
		\includegraphics[width=0.75\columnwidth]{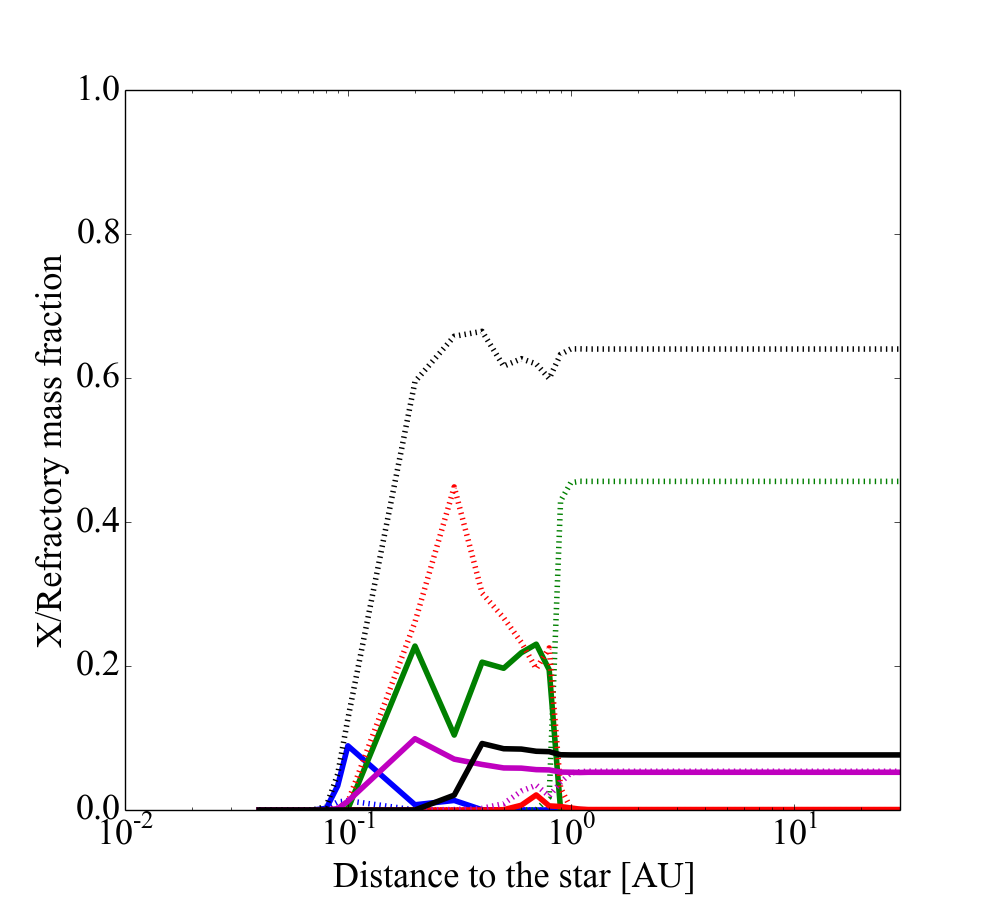} \\
		\caption{Distribution of silicates relative to all condensed refractory molecules in one of the simulated discs ($\Sigma_{0}$=95.844 g.cm$^{-2}$, a$_{core}$=46 AU and $\gamma$=0.9) for the model "without" organic compounds. Full blue: CaO*Al$_2$O$_3$*2SiO$_2$; dotted blue: *2CaO*Al$_2$O$_3$*SiO$_2$; full green: Mg$_2$SiO$_4$; dotted green: Mg$_3$Si$_2$O$_5$(OH)$_4$; full red: Fe$_2$SiO$_4$; dotted red: MgSiO$_3$; full purple: CaMgSi$_2$O$_6$; dotted purple: FeSiO$_3$; full black: NaAlSi$_3$O$_8$; dotted black: All silicates.} 
		\label{fam_vs_A}
	\end{figure}

			Based on \cite{Lodders2003}, the C/O ratio of the solar nebula is around 0.5. Thus, Si forms mainly silicates in the solar system, which is verified by the calculations that show that carbides make up less than 1\% in mass of solids accreted by the final planets (see Figure \ref{fam_vs_A}). The mass fraction of C relative to the mass of solids shown in Figure \ref{C_vs_M_vs_A} is thus due to refractory organic material and volatile compounds for population B and are only due to refractory organic material for population A, as C is incorporated only in volatile and refractory organic compounds. \\
			
		\subsection{Silicates} \label{sect_silic}
\begin{figure} \centering
		\includegraphics[width=0.75\columnwidth]{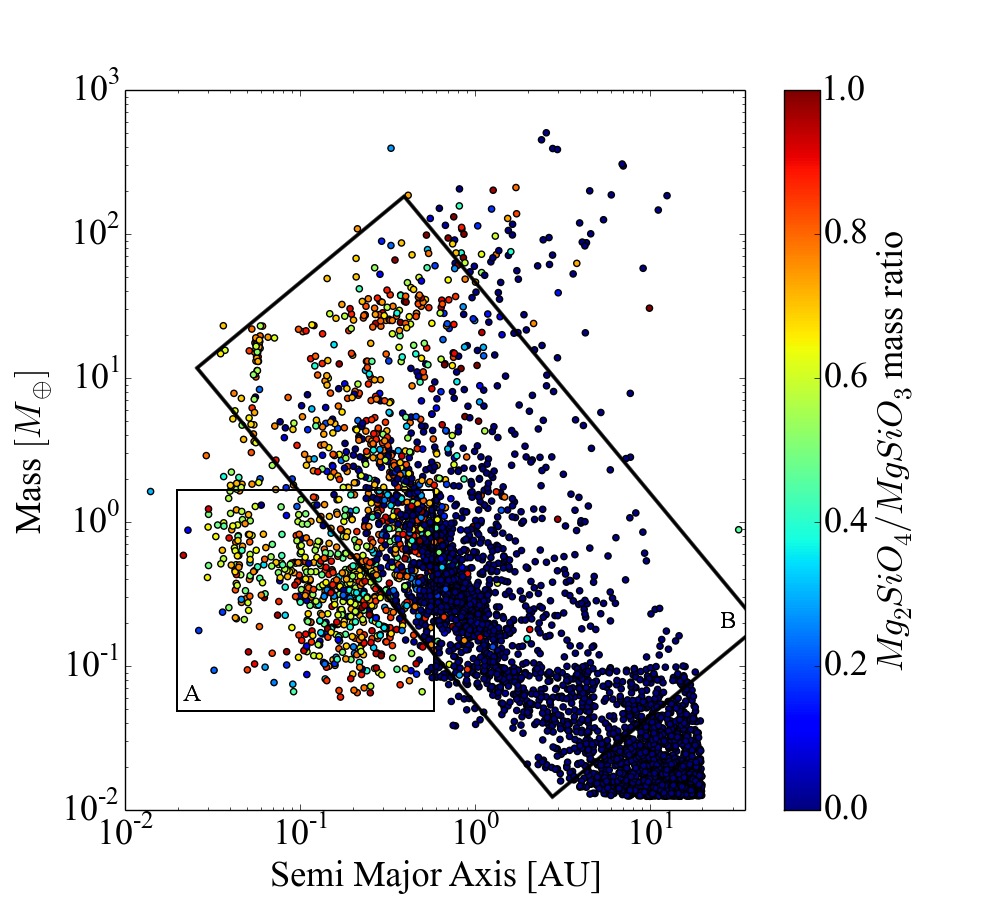} \\
		\caption{Mg$_{2}$SiO$_{4}$/MgSiO$_{3}$ ratio. The figure is identical for the model "with" refractory organic compounds.} 
		\label{MgSiO_vs_M_vs_A}
	\end{figure}

			Silicates are an important family of minerals in rocky planets where they make up the major part of the mantle and crust of planets, as it is the case for  the Earth, Mars, and Venus \citep{Morgan1980}. Their distribution is ruled by the Mg/Si elemental ratio of the host star \citep{DelgadoMena2010}. For Mg/Si values lower than 1, Mg forms orthopyroxene (MgSiO$_{3}$) and the remaining Si forms other silicate minerals like feldspars (CaAl$_{2}$Si$_{2}$O$_{8}$, NaAlSi$_{3}$O$_{8}$) or olivine (Mg$_{2}$SiO$_{4}$). Feldspars and olivine are therefore expected to be much less abundant than pyroxene; thus, the ratio between O in silicates and Si is expected to be close to 3. 
			For Mg/Si ratios ranging from 1 to 2, Mg is distributed between pyroxene and olivine. For Mg/Si higher than 2, Si forms olivine, and the remaining Mg forms other phases, mostly oxides. The Mg/Si ratio of the Sun is 1.02$\pm$0.10 \citep{Lodders2003}. It is thus expected that the rocky planets in the solar system are dominated by early formation of pyroxene with lesser quantities of olivine.
			Figure \ref{MgSiO_vs_M_vs_A} shows the ratio between olivine and pyroxene abundances. As one can see, this ratio varies from 0 to 1 with a tendency for values between 0 and 0.5 for rocky planets and close to zero for population B. \\

			Pyroxene (MgSiO$_{3}$) and olivine (Mg$_{2}$SiO$_{4}$) are not the only silicates to form. Figure \ref{fam_vs_A} shows the total fraction of silicates relative to all refractory phases formed as a function of the distance to the star in one of the simulated discs. At radii smaller than 1 AU, Mg-pyroxene and -olivine are the major species and can form up to about 65\% of all refractory material. They indeed condense at higher temperatures than the other species and only small amounts of less refractory phases are formed due to the high temperatures. \\
			
			The mass fraction of silicates relative to all refractory phases decreases a little from  at 0.8 AU and stays constant beyond 1 AU. In these calculations, the temperature at 1 AU is 230 K. Thus, the refractory composition  from this position or this temperature does not change significantly to generate important variations in this fraction and stays constant at $\sim$47 wt \%.  %However, we cannot differentiate between the crystalline silicates from the amorphous silicates which are coexisting (\cite{Colangeli2004}).

	\section{Influence of the irradiation}
		The model used in these simulations neglects the irradiation coming from the star. The temperature and pressure profiles of the discs can then be underestimated if this irradiation is not taken into account \citep{Garaud2007}. The disc produced with the irradiation will be warmer \citep{Fouchet2012}, which causes the composition of the planetesimals to be different. Consequently, a depletion in volatile molecules is expected, shifting the position of the iceline outwards at the same time. \\
		
		Two additional simulations were run: 1. To test the influence of the irradiation on the chemical composition, the same planet formation model was used. Only the composition of the planetesimals (and therefore in the composition of planets) was changed according to the presence of the irradiation. 2. To test the influence of the irradiation on the planet formation, the planet formation model and the chemical model were changed to take into account the irradiation.

		\subsection{Influence on the chemical abundance}
\begin{figure} \centering
			\includegraphics[width=0.75\columnwidth]{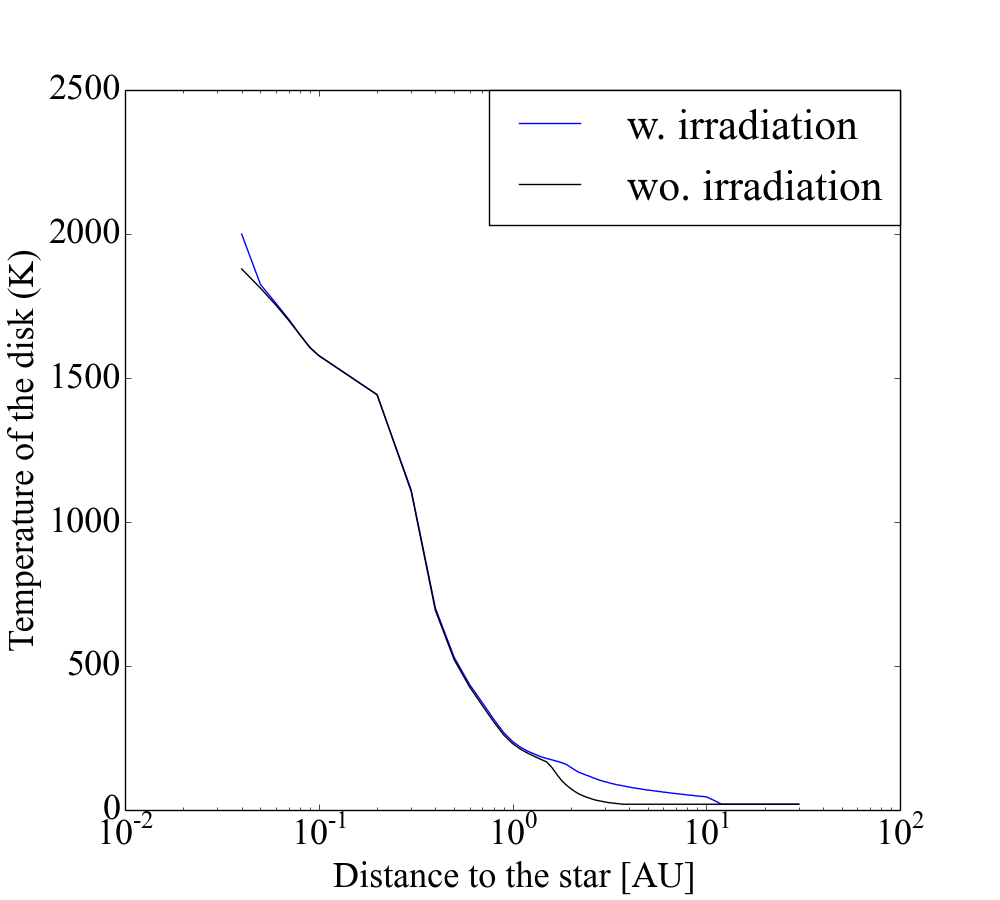} \\
			\caption{Temperature profiles of a disc (a$_{core}$ = 46 AU, $\Sigma_0$ = 213.86 g.cm$^{-2}$, $\gamma$ = 0.9) for irradiated and non-irradiated calculations. Both temperature profiles are very similar and the changes are of around 100K between 2 and 10 AU only.} 
			\label{temp_profile}
		\end{figure}

\begin{figure} \centering
			\includegraphics[width=0.75\columnwidth]{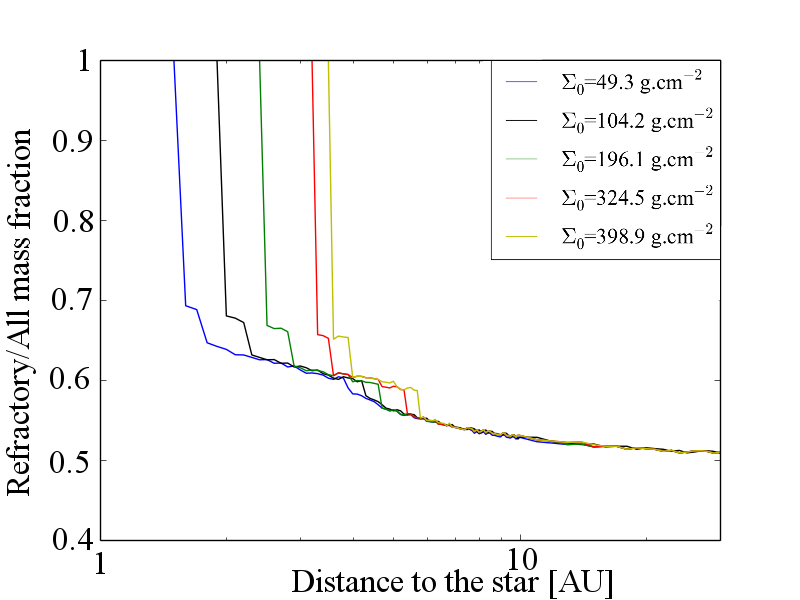} \\
			\caption{Refractory mass fraction relative to all condensed molecules as a function of the distance to the star in five discs of different masses (a$_{core}$ = 14 AU and $\gamma$ = 0.8, fixed) in an irradiated disc. The position where the fraction drops is the position of the iceline.} 
			\label{disk_w}
		\end{figure}

				The simulation was run with the model without organic compounds, which best reproduces the observations in a non-irradiated disc. The temperature profiles of one of the discs for both cases is shown Figure \ref{temp_profile}. The temperature profiles are very similar; thus, the final composition of planets are expected to be similar to the case of a non-irradiated disc for planets that formed at low or high distances to the star.  As shown in Figure \ref{disk_w}, this position is, however, not shifted significantly; the maximum shift being of around 0.5 to 1 AU outwards. However, the shape of the curve is changed with a slower convergence to the smallest values. 
		

\begin{figure} \centering
			\includegraphics[width=0.75\columnwidth]{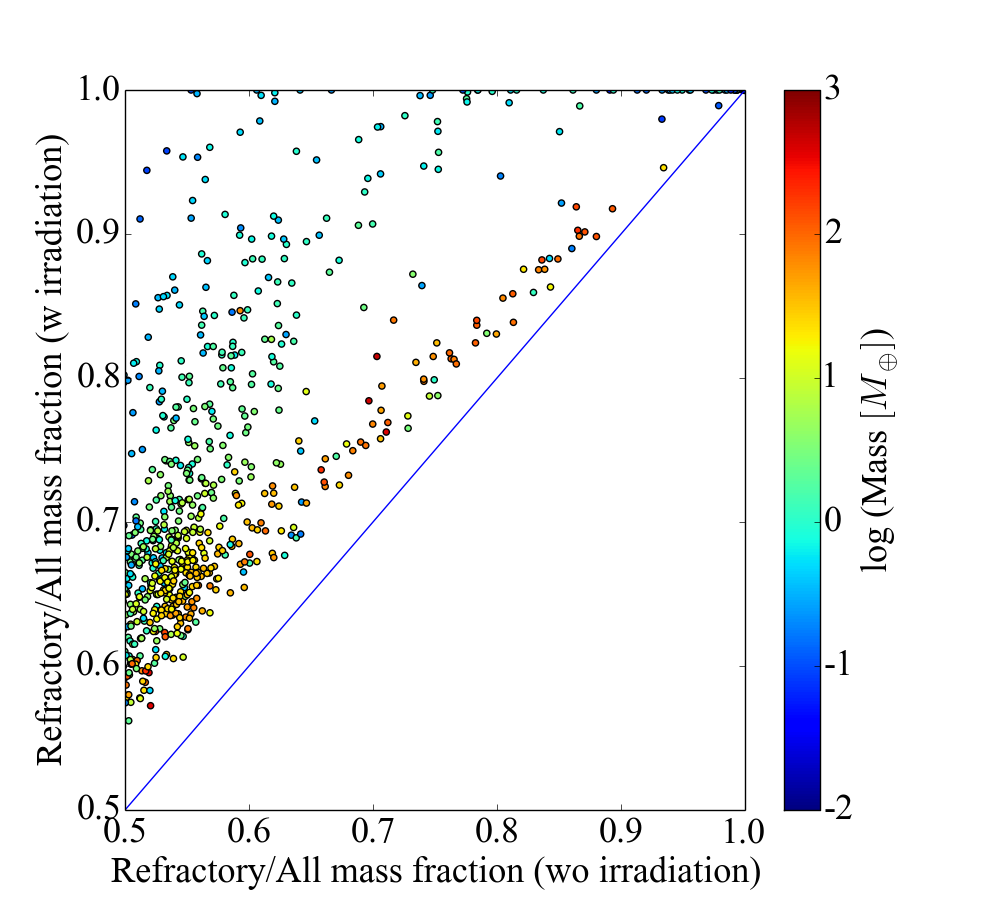} \\
			\caption{Refractory mass fraction relative to all condensed molecules in the irradiated case versus the non-irradiated case. The color bar indicates the logarithm of the mass in Earth masses. } 
			\label{ref_irr}
		\end{figure}

		It is expected that the giant planets that formed mainly at a large semi-major axis are the planets for which the compositions change the least. Figure \ref{ref_irr} shows the influence of the irradiation in the expected enrichment in refractory components. Because the disk is warmer, planets will contain more refractory material. Giant planets are indeed the planets with the least enrichment in refractory material with 1.2 times more refractory elements than in the non-irradiated case, whereas small inner planets are highly enriched. 
		
\begin{figure} \centering
			\includegraphics[width=0.75\columnwidth]{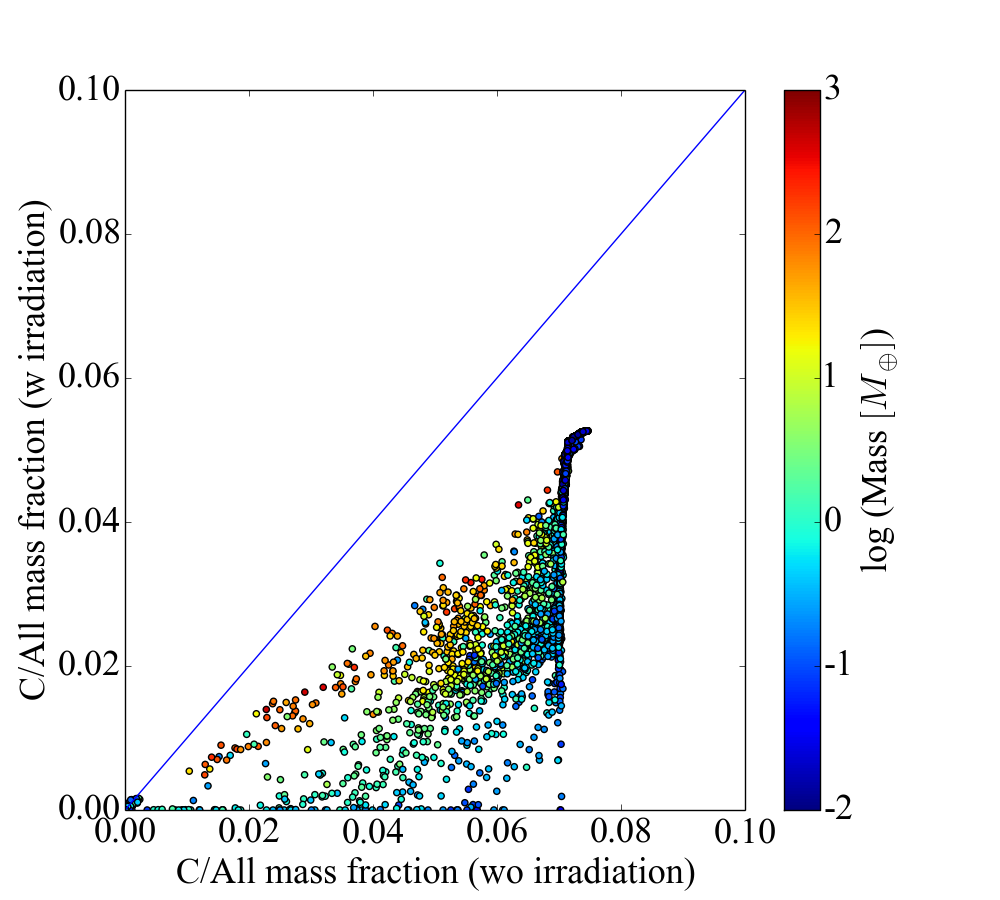} \\
			\caption{Same as Figure \ref{ref_irr} but for carbon. } 
			\label{C_irr}
		\end{figure}

\begin{figure} \centering
			\includegraphics[width=0.75\columnwidth]{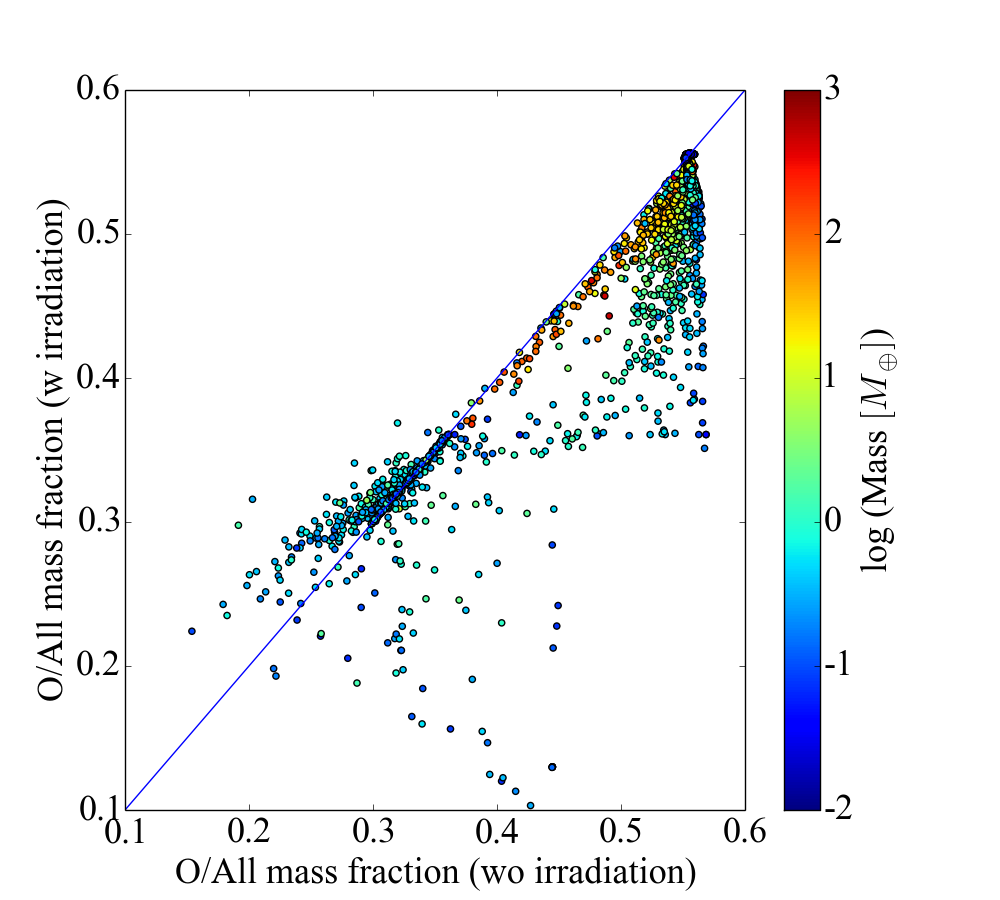} \\
			\caption{Same as Figure \ref{ref_irr} but for oxygen.} 
			\label{O_irr}
		\end{figure}

\begin{figure} \centering
			\includegraphics[width=0.75\columnwidth]{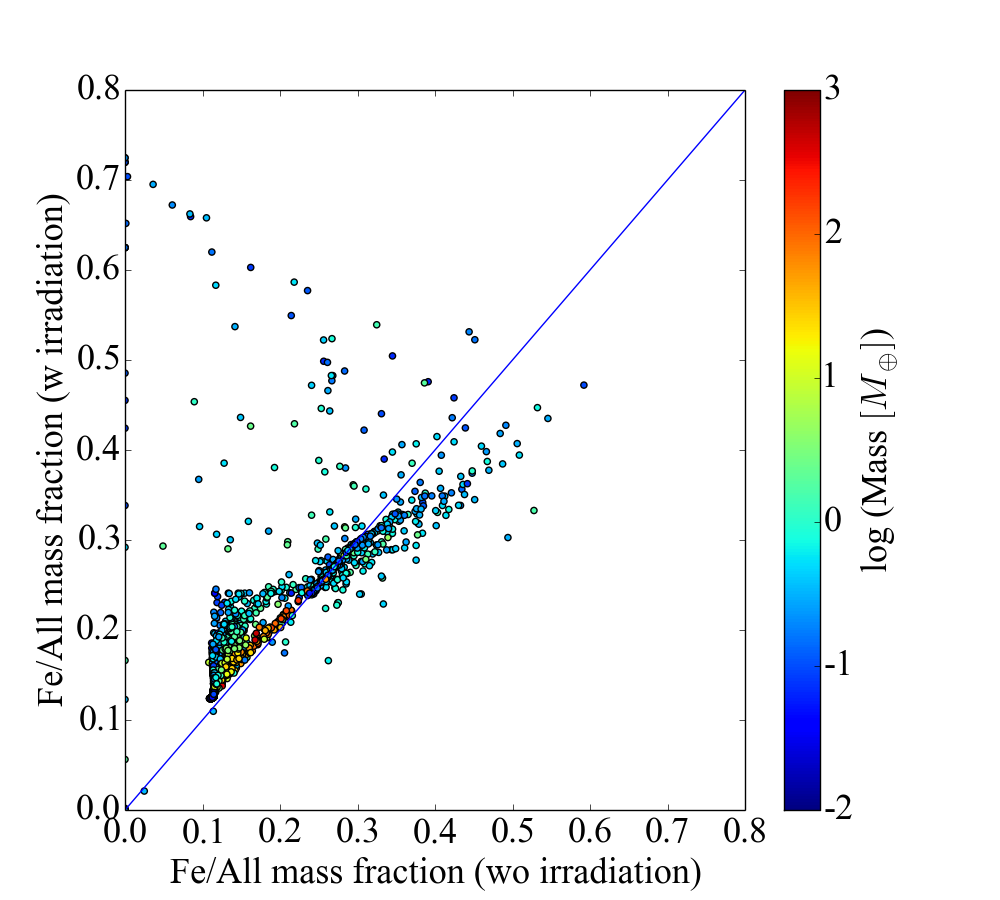} \\
			\caption{Same as Figure \ref{ref_irr} but for iron.} 
			\label{Fe_irr}
		\end{figure}

\begin{figure} \centering
			\includegraphics[width=0.75\columnwidth]{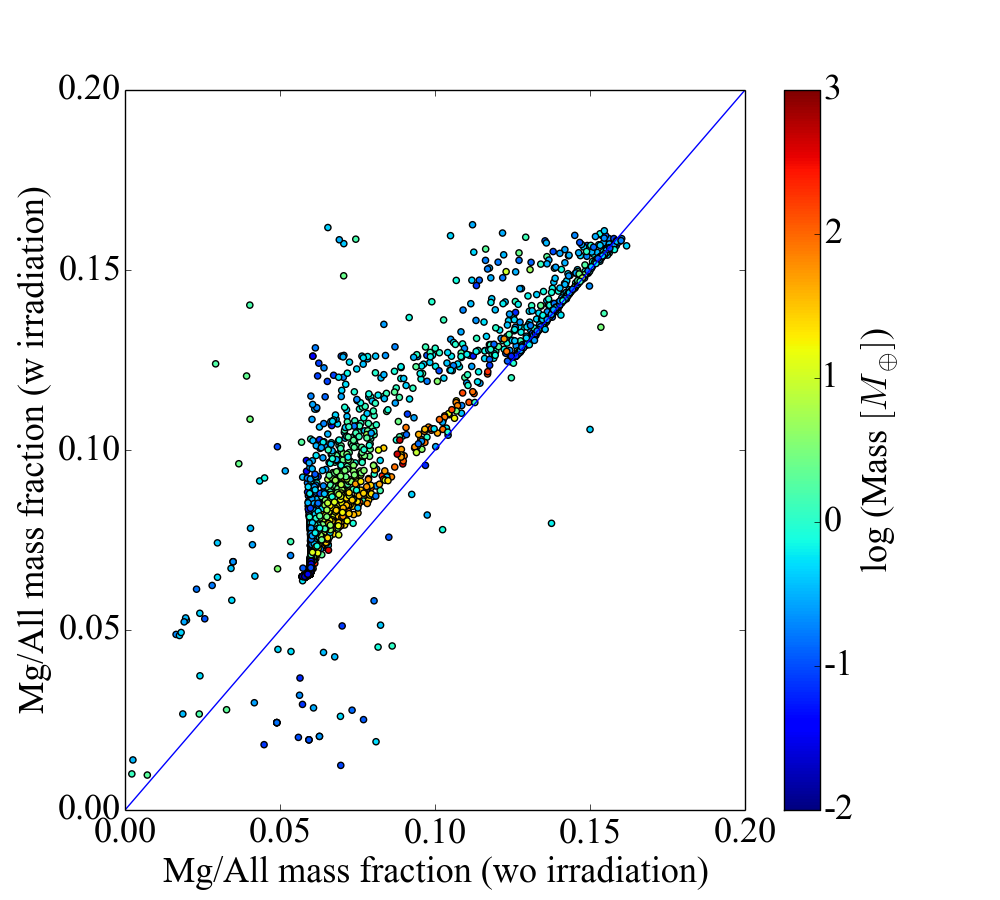} \\
			\caption{Same as Figure \ref{ref_irr} but for magnesium.} 
			\label{Mg_irr}
		\end{figure}

\begin{figure} \centering
			\includegraphics[width=0.75\columnwidth]{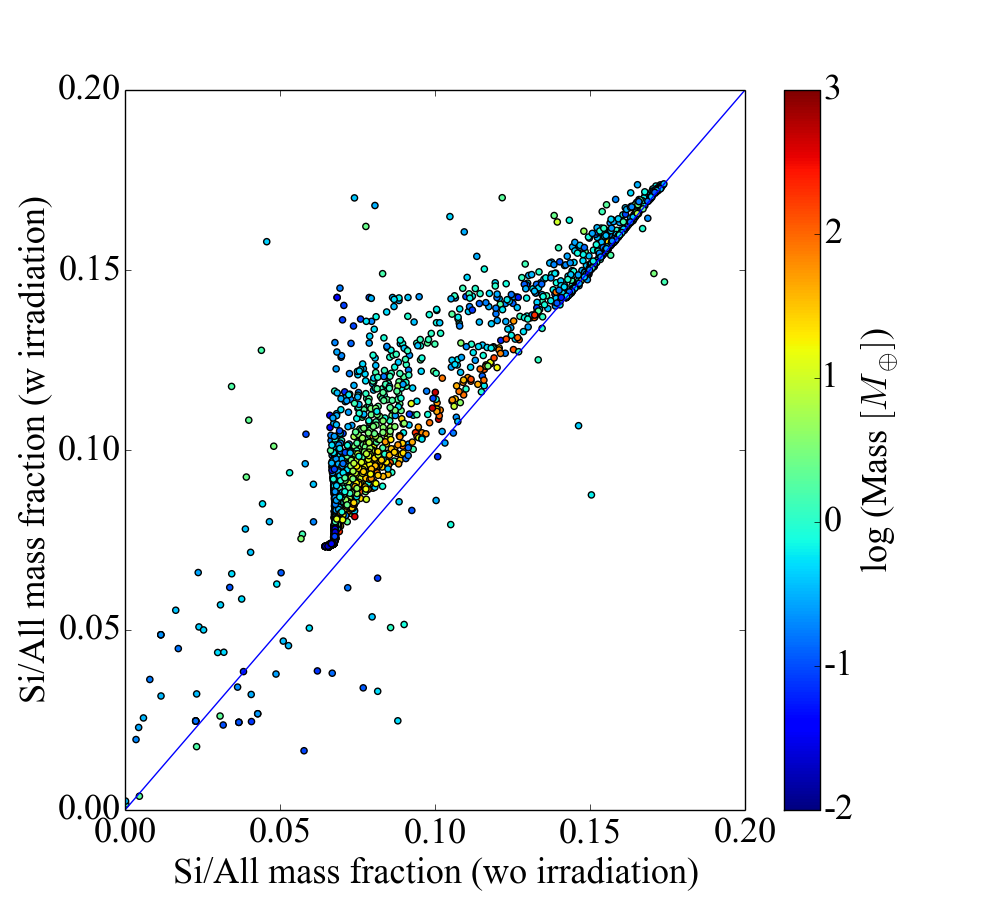} \\
			\caption{Same as Figure \ref{ref_irr} but for silicon. } 
			\label{Si_irr}
		\end{figure}

\begin{figure} \centering
			\includegraphics[width=0.75\columnwidth]{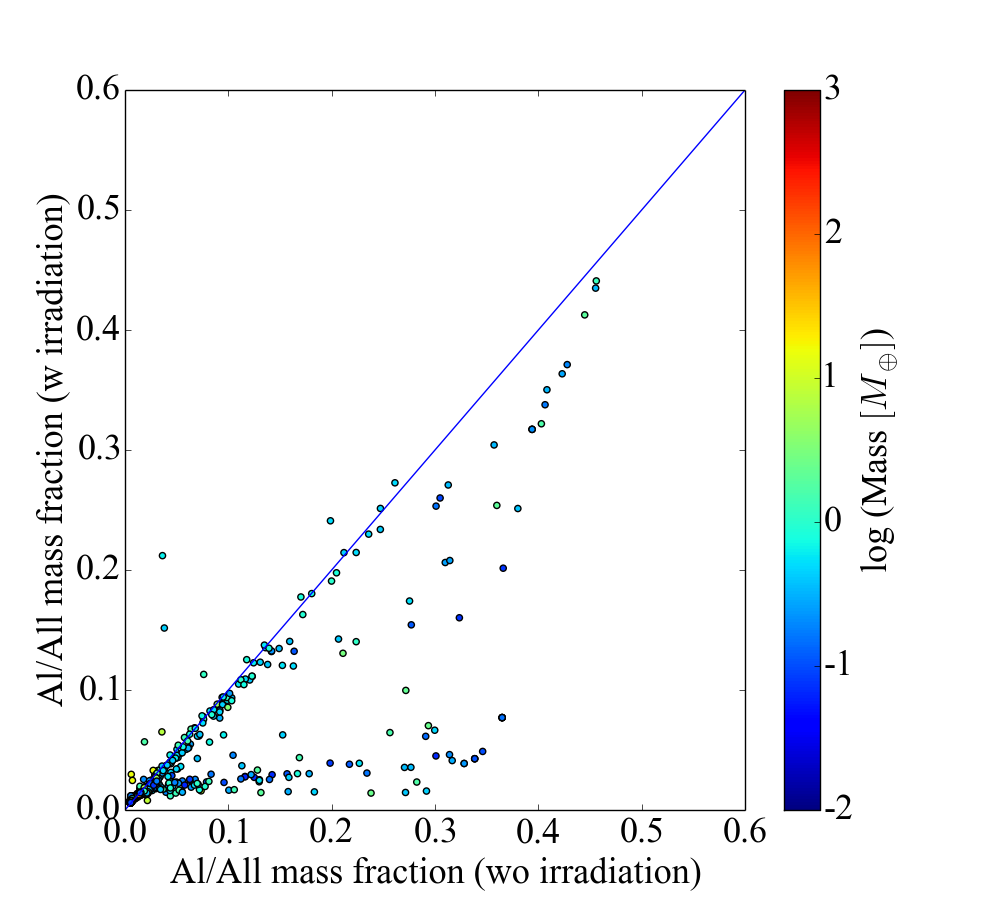} \\
			\caption{Same as Figure \ref{ref_irr} but for aluminium.} 
			\label{Al_irr}
		\end{figure}

\begin{figure} \centering
			\includegraphics[width=0.75\columnwidth]{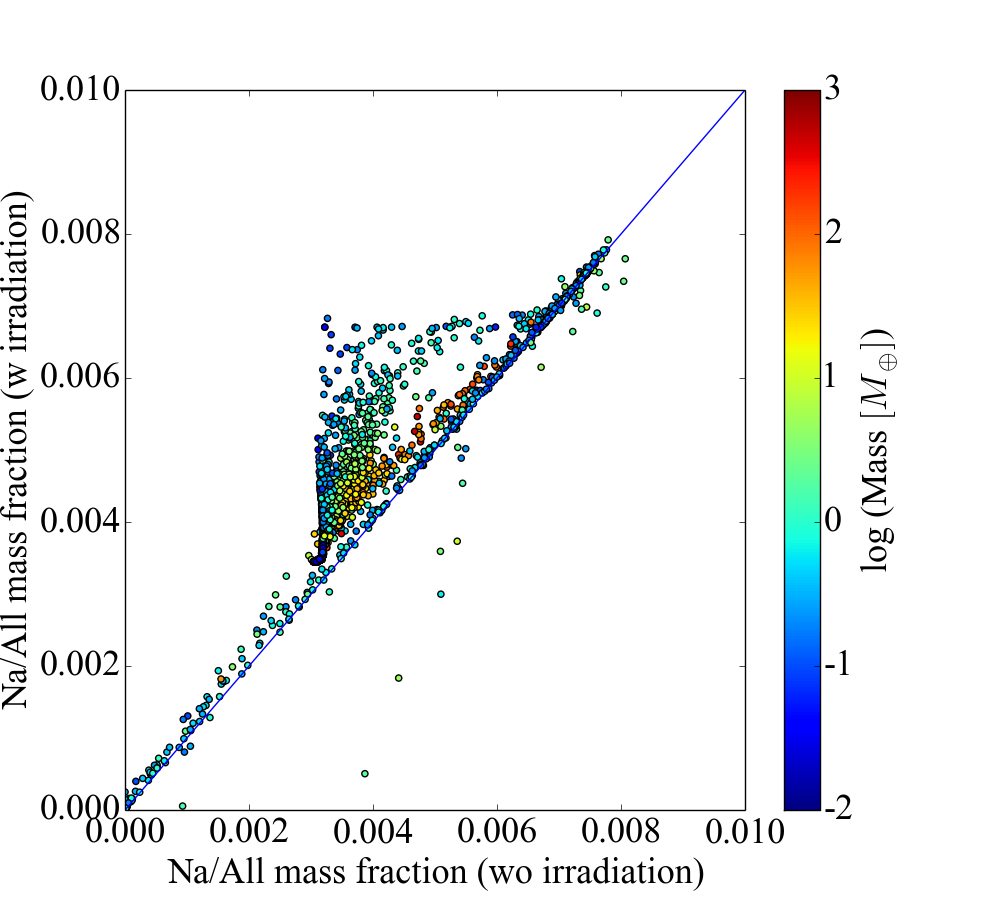} \\
			\caption{Same as Figure \ref{ref_irr} but for sodium. } 
			\label{Na_irr}
		\end{figure}

\begin{figure} \centering
			\includegraphics[width=0.75\columnwidth]{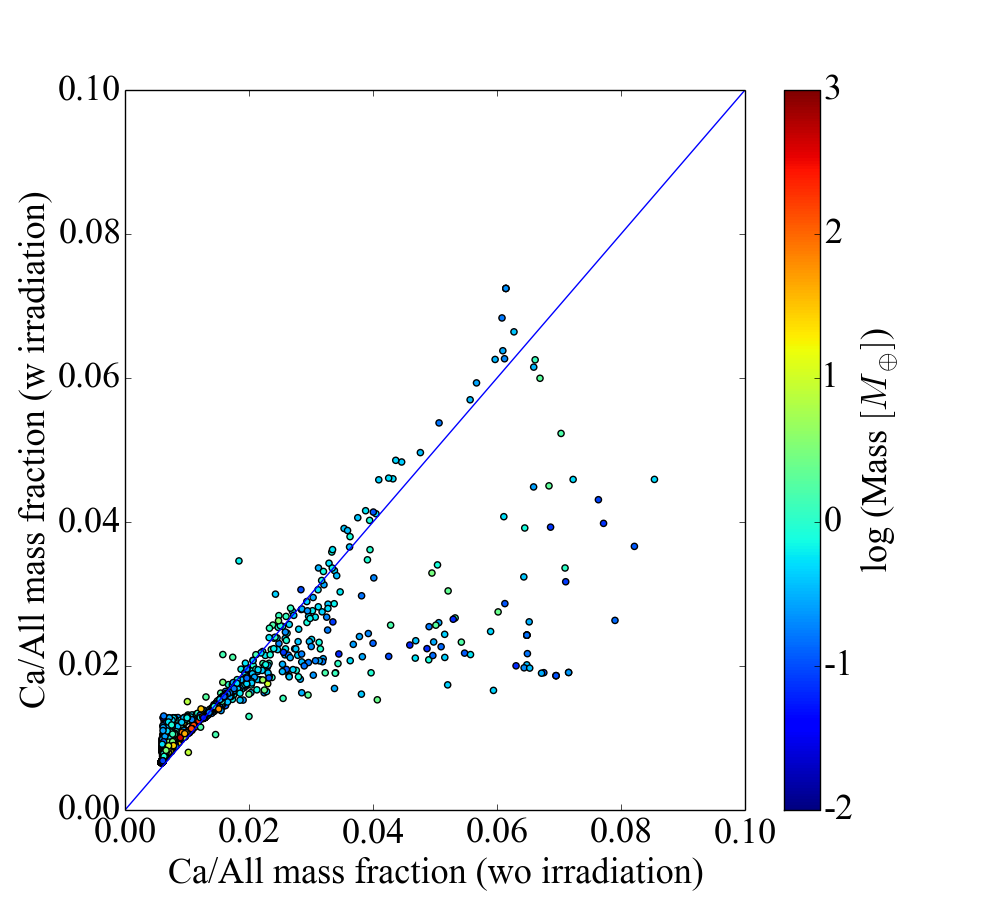} \\
			\caption{Same as Figure \ref{ref_irr} but for calcium. } 
			\label{Ca_irr}
		\end{figure}

		This pattern can be seen in the abundance of every element. Figures \ref{C_irr} to \ref{Ca_irr} show the influence of the irradiation in the abundance of the metallic elements considered in this study.
		For giant planets,  the relative abundances do not vary a lot, showing the very limited influence of the irradiation on planets that mainly formed beyond the iceline. The influence of the irradiation is, however, stronger for the other planets. The abundance of carbon is lower but does not vary enough to be of influence for the final composition of planets (the same can be observed for Na and Ca). Due to the outward shift of the iceline, the abundance of O is lower for most of the planets, except for those that formed mainly between the star and the iceline as in both cases and for whose O abundance is increased by 10 wt \% at most. The same phenomena can be observed for Fe, Mg, and Si. The results are very similar but due to the shift of the iceline; the amount of these 3 elements is higher than before with an interesting barrier  at 30 wt \% for Fe and at around 16 wt \% for Mg and Si. This could indicate a process similar to the radial mixing of planetesimals as modeled by \cite{Lewis1988}.
		For most of the planets, the Al abundance is lowered and reproduces better what is observed in our Solar System. 
		
		%\clearpage
		
		\subsection{Influence on planet formation}
			The formation of planets and their final composition depends on the presence of irradiation. The  model of planet population synthesis of \cite{Alibert2005d} was used in \cite{Fouchet2012} to reveal the effects of the irradiation on planet formation. In a classical way such that a gap is opened if $R_h$>$H_{disk}$ (with $R_h$ the Hills radius of the planet and $H_{disk}$ the height scale of the disk), the non-irradiated case and the irradiated case differed in the formation of giant planets. In the model with irradiation, proto-planets could not undergo a type II migration, resulting in the absence of giant planets in the a-M diagram. As expected, the irradiation changes $H_{disk}$, resulting in the possibility for the planets to undergo a type II migration later than without irradiation.
			 The study conducted by \cite{Fouchet2012} showed, however, that this population of giant planets can be recreated if the more realistic gap opening criterion of \cite{Crida2006} is taken. Consequently, the difference in both population of planets (irradiated and non-irradiated) is only small. We use this approach to include irradiation in our previous model. More details for this model are given in \cite{Fouchet2012}.\\
			 
\begin{figure} \centering
				\includegraphics[width=0.75\columnwidth]{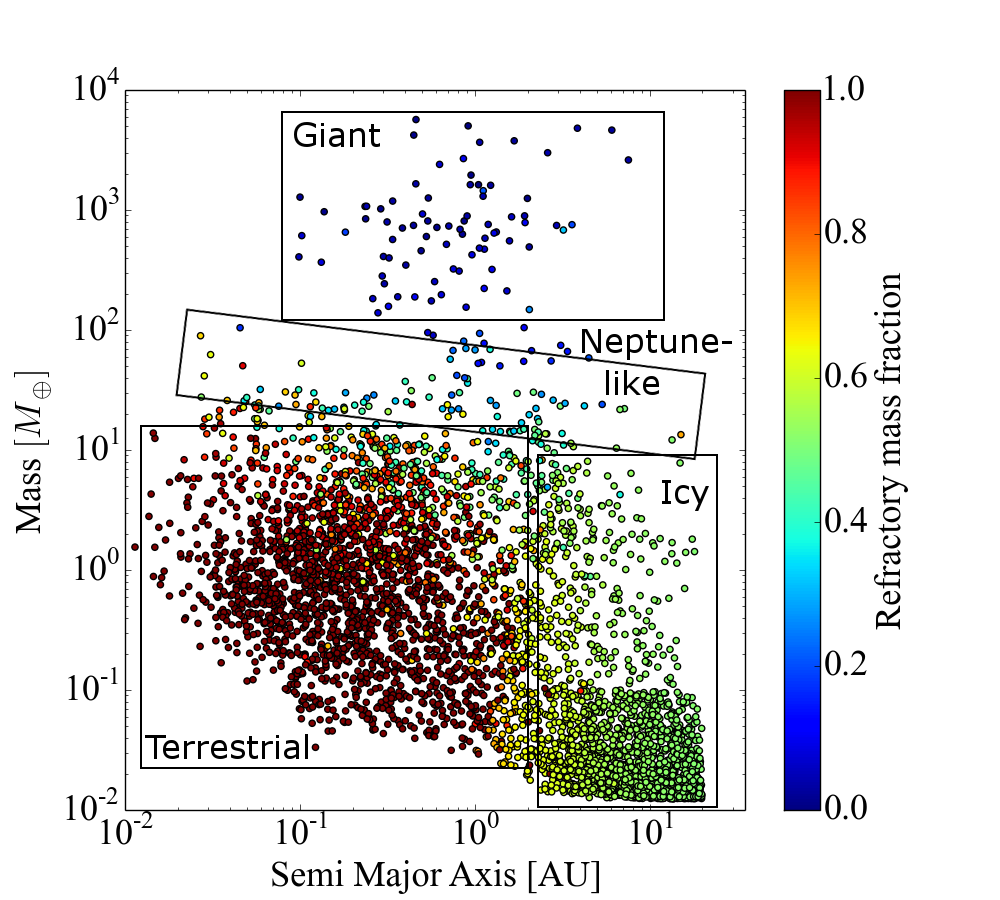} \\
				\caption{Refractory mass fraction relative to the total mass of the planet (solids + gas) for a fully irradiated model.} 
				\label{pop_irr}
			\end{figure}

			 Figure \ref{pop_irr} shows the refractory mass fraction of the new population. This was computed using the same initial conditions as the previous population (disc mass, lifetime of the disc...), and the only changes are made to the irradiation. The results show that both irradiated and non-irradiated (Figure \ref{population}) populations are similar. The major change comes from the amount of refractory elements in the planets due to the irradiation. As discussed in the previous section, the iceline moves outwards, resulting in an enrichment in refractory elements. An interesting point is that it is possible to form more Earth-like planets at the current location and the current mass of Earth in the Solar System than in the non-irradiated case with this model. \\ 
			 
			 It should be kept in mind that our model of irradiation assumes that the flaring angle of the disk is constant and equal to its equilibrium value \citep[see][]{Hueso2005, Fouchet2012}. Moreover, the disc is assumed to be totally irradiated, namely that no shadowing effect can take place in the disc. These are obvious simplifications, and the model, which is presented in this section, probably overestimates the effect of the irradiation. We therefore suggest that a more realistic model should lie between the non-irradiated and the fully irradiated model.		 
		%En terme de composition, cela ne bouge que trs peu, comme le montre les figures (blablabla2) ˆ (blablabla3). les ŽlŽments les moins abondants dans les simulations non irradiŽes sont toujours aussi peu abondants, voire absents si une irradiation totale est prise en compte. Ainsi, la quantitŽ d'aluminium prŽsent dans les plantes irradiŽes a chutŽ pour n'tre presque plus prŽsent. C'est le seul ŽlŽment dont l'abondance est aussi fortement changŽe, nŽanmoins c'est l'un des ŽlŽments les moins abondants dans les deux cas (irradiŽs et non irradiŽs) et influe peu sur la plante en elle-mme. Comme attendu, la proposition d'ŽlŽments rŽfractaires augmente d'environ 20\%, diminuant de mme la proportion de volatiles dans ces plantes. Par ailleurs, les plantes les moins touchŽes par l'irradiations sont les plantes gŽantes, formŽe en grande partie dans les rŽgions les plus externes du disque dans lesquels les cas irradiŽs et non irradiŽ ont un profil de tempŽrature extrmement similaire

	\section{Discussions} \label{discussions}
	
		\subsection{Importance of the planet formation model}
		
		Currently, the formation of planets from a stellar nebula can only be understood through modeling, since the process cannot be observed directly and direct information on the composition of the resulting planets is only available for some solar system objects. Thus, the diversity of planets, their formation paths and their physical and chemical models are essential to evaluate the possible scenarios. These models show that the disc parameters have a major influence on the planet formation pattern and on the size and chemical composition of the formed planet (see Figure \ref{pos_iceline}). If we consider two planets formed at 2 AU with the same final mass in two discs of different masses, which are given values such as $\Sigma_0$ = 49.3 g.cm$^{-2}$ and  $\Sigma_0$ = 196.1 g.cm$^{-2}$ of Figure \ref{pos_iceline}, the amount of refractory material and chemical composition of both planet may be very different (55 wt \% and 100 wt \% respectively). \\
				
\begin{figure} \centering
			\includegraphics[width=0.75\columnwidth]{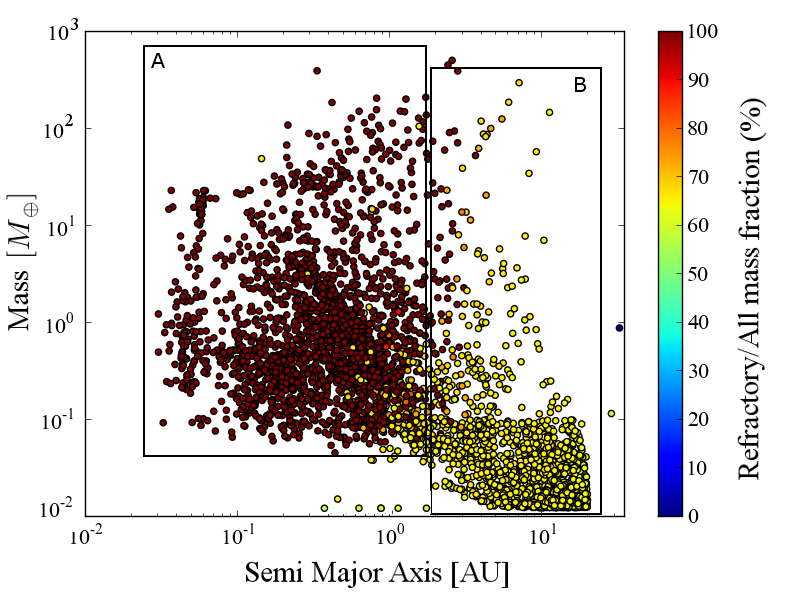}
			\caption{Refractory mass fraction relative to all condensed molecules  as a function of the mass of solids accreted by the planet and the final position of the planets in the case of \textit{in situ} formation for model "with".}  
			\label{refracto_nonself}
		\end{figure}

		Self-consistency in the planetary formation model is demonstrated to be of upmost importance. Figure \ref{refracto_nonself} shows the refractory mass fraction of planets that formed \textit{in situ}; that is the planets are formed at their final distance from the central star without moving radially. In this case this non-self-consistent model shows that the chemical composition of the giant planets, which formed closer than 1 AU from the central star, is not consistent with the chemical composition observed in the gas giants. Comparing the results obtained in Figure \ref{refracto2_vs_M_vs_A} with the results of Figure \ref{refracto_nonself}, both populations A (containing more than 90 wt \% of refractory component) and B (the other planets) are very different. Population A has expanded in both mass and semi-major axis range and population B mainly consists of planets that did not evolve much. This highlights the need of a self-consistent planet formation model.
		
			\subsection{Equilibrium} \label{disc_ab_cn}
		     	The equilibrium assumption is something that cannot be verified in systems other than the Solar System. Chemical reactions could still be occurring, while the planetesimals are forming and cosmic rays or UV irradiation could lead to a non-equilibrium state. {However, non-equilibrium chemistry in discs is difficult to constraint because of the multitude of chemical reactions that potentially can occur and that require exact knowledge of chemical properties of all species and compounds involved, which is currently not possible. The possibility of remnants from the interstellar medium, especially for highly refractory elements such as Ca or Ti, is not taken into account, as their abundance is low enough to be neglected.} %Equilibrium could thus not be verified for other planetary systems and calculations under non-equilibrium state will be needed. 
			\\
			
			The assumption about equilibrium is directly linked to the assumption that the pressure does not vary significantly from the mean values.  The use of an average pressure is studied using the standard deviation to the mean pressure values. The pressure boundaries can vary by one order of magnitude; therefore, more calculations for these boundaries were run. The results show that the use of a mean pressure to simplify the calculations is valid with less than 2\% deviation from the values of the chemical compositions of the planets. The calculations are thus not highly dependent on the pressure of the system, within at least the boundaries encountered in our models.
				
		\subsection{Refractory Organic Compounds}
\begin{figure} \centering
			\includegraphics[width=0.75\columnwidth]{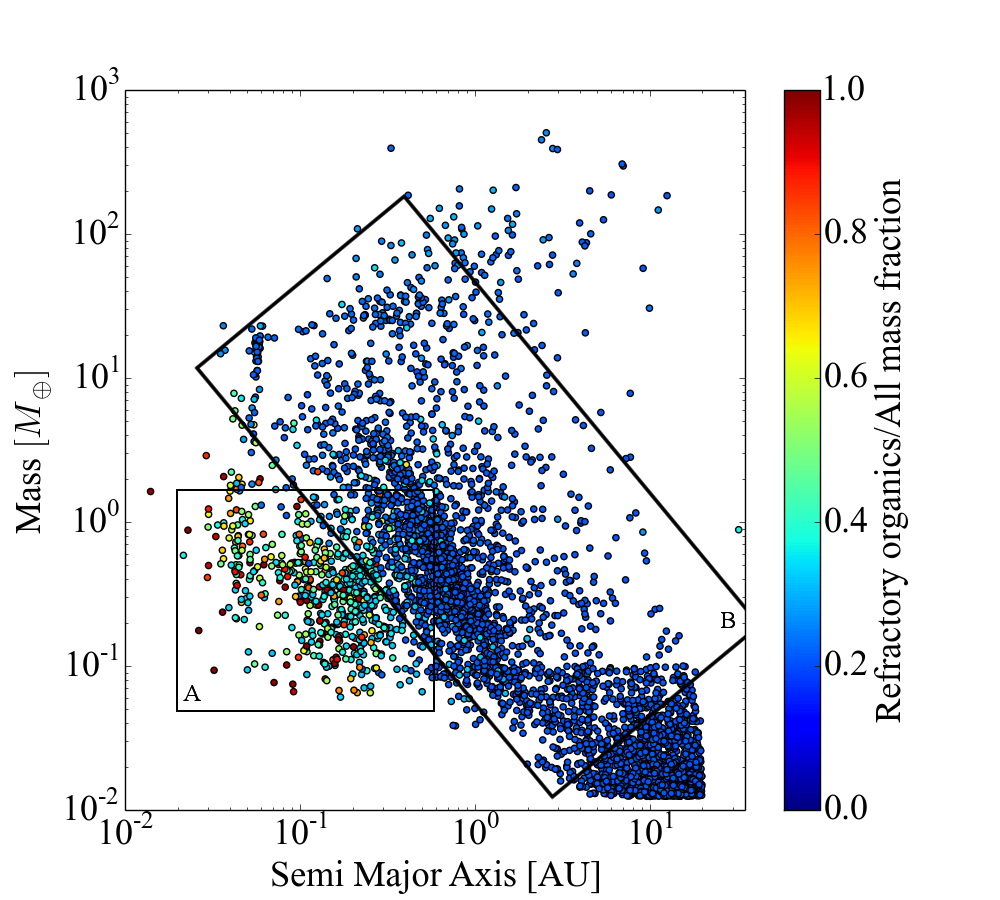}
			\caption{Refractory organic compounds mass fraction relative to all condensed molecules  as a function of the mass of solids accreted by the planet and the final position of the planets in the model "with".}  
			\label{organiques}
		\end{figure}

			The presence of refractory organic compounds in the solar nebula still has to be confirmed. Although evidence of such compounds has been found in proto-stars (band at 6 $\mu$m, see \cite{Gibb2002}) and also in observations and measurements of comets \citep[see][for a review]{Cottin1999}, the nature of  the major molecules remains unknown. In the calculations presented in this paper, two extreme cases were considered. The model "with" assumes that refractory organic compounds do not react with the other compounds and  that they are also not destroyed by any process. The model "without" considers that no organic compounds formed. Figure \ref{organiques} shows the mass fraction of refractory organic compounds in the model "with". In planets of population B, refractory organic material represents 10 wt \% of the solids accreted by the planets, while it varies between 40 wt \% and 100 wt \%  for planets of population A. Planets with such high fractions of refractory organic material are planets that are very close to the star and that have evolved mainly between 0.04 AU and 0.3 AU in relatively massive discs. It implies that refractory inorganic compounds could not condense, and organics then become the most abundant component of these planets, as discussed in Sect. \ref{planet_compo}. Such planets are unlikely to exist in the Universe. However, calculations for less organic compound rich planets enable us to find an upper limit for the abundance of organic compounds in planets: 40 wt \% of solids accreted and 30 wt \% of the total mass of the planet (solids + H$_2$).
			
			Recent studies toward pre-stellar cores \citep{Bacmann2012} suggest that complex organic compounds are already formed when the molecular cloud collapses to form a star. \cite{Cottin1999} also showed that refractory organic material can be formed before the appearance of the disc.  Moreover, refractory organic compounds are thought to condense at similar temperatures as inorganic compounds. However, their fraction seems small compared to the other species present at the same time, thus signifying that most of these organic compounds form after the appearance of the disc, which is either during the process of planet formation or once the disc has disappeared. We cannot, however, determine which is correct. Although the present study suggests that they are not formed at all because the results of the model "without" are closer to what is observed in the Solar System, we cannot exclude the possibility of a formation during the process of planet formation.
					
		\subsection{Migration of planetesimals} \label{migration_plan}
			While the possibility of ongoing chemical reactions has been discussed in this contribution, the radial drift of the planetesimals due to the effect of gas drag and gravitational torques was not yet considered. The migration of planetesimals is potentially an important process during planet formation, as it may lead  to depleted regions or deliver more volatile material to the inner part of the disc from beyond the iceline.
			However, the assumption of no radial drift of planetesimals in this study remains valid for larger planetesimals (~1 km), if they are formed rapidly. If one of these criteria is not valid, radial drift needs to be taken into account. This will be done in a future study.
			{It can be discussed that pre-accretionnary drifting may play a role. However, this will highly depend on the planetesimals formation process, which is still a matter of research. This contribution assumes that planetesimals form rapidly enough for this process to be neglected.}
			
	\section{Summary and conclusion} \label{ccl}
		In this contribution, we have presented the results of calculations to determine the refractory composition of planetesimals and planets by combining the planet formation model of \citet{Alibert2005d, Alibert2013}  with a chemical model, where compositions are ruled by equilibrium condensation in a self-consistent way. The main results of the study are as follows
		
		\begin{itemize}
			\item Using the planet formation in a self-consistent way, it has been shown that  a large diversity of planets can be formed when starting with different disc parameters but identical chemical bulk compositions of the solar nebula
			\item Planets formed can be either terrestrial (i.e, rocky) planets (19\%), icy planets (66\%), or gas giants (15\%).
			\item Rocky planets contain less than 1wt \% of volatile compounds.
			\item Gas giants and icy planets are composed of $\sim$55 wt \% of refractory material (respectively $\sim$65 wt \%) for a model "without" (respectively a model "with") refractory organic compounds. In a case of a stellar nebula of solar composition, refractory organic compounds can represent up to 40 wt \% of solids accreted by a planet.
			\item Silicates dominate the refractory composition of the discs at every distance to the star with a concentration up to 65 wt \% for the planetesimals in the inner part of the disc and 45 wt \% in the outer parts. Major silicate minerals species are MgSiO$_3$, Mg$_2$SiO$_4$, and Mg$_3$Si$_2$O$_5$(OH)$_4$.
			\item  The two models described in this study give similar results for rocky planets, but the absence of refractory organic compounds of the refractory system (model "without" ) gives results that are closer to what is observed for the Solar System. %This suggest that refractory organic molecules are not present in high abundances in the disc.
			\item Refractory material represents from 50\% to 100\% of the mass of solids accreted by a planet that has an ice/rock ratio of up to 1, which is in good agreement with other models but which is smaller than the value used in recent planet formation models. This ratio also depends on the position of the iceline. 
			\item We find similar results as \cite{Lewis1972} for Mercury-like planets with 70 wt \% of Fe. However, the calculations do not consider the radial drift of planetesimals so far, which could lead to a smaller abundance by a factor of 2.
			\item Different discs are required to explain the diversity of the formed planets. Using the planet formation model in a self-consistent way is also needed.
			\item {Irradiation moves the position of the iceline outwards. Consequently, the planets formed are richer in refractory elements.}
		\end{itemize}		
		
		The results of this study are in good agreement with cometary observations with chemical abundances known for CI chondrites and the solar photosphere \cite{DelgadoMena2010}. They are also consistent with the previous work of \cite{Elser2012}. 
		In a future study, we will extend the model to non-solar composition, include the radial drift of planetesimals, and compute the radii of planets in a self-consistent way using the results from this study.
		
		\paragraph{{Acknowledgments}} This work was supported by the European Research Council under grant 239605, the Swiss National Science Foundation, and the Center for Space and Habitability of the University of Bern.
	
	\bibliographystyle{aa}
	\bibliography{biblio}

\end{document}